\documentclass{aastex62}

\usepackage{comment}
\usepackage{amsfonts}
\usepackage{amsmath}

\mathchardef\mhyphen="2D

\graphicspath{{./}{figures/}}

\received{\today}
\revised{\today}
\accepted{\today}
\submitjournal{ApJ}

\shorttitle{GRB model comparison}
\shortauthors{Acuner et al.}

\begin{document}

\title{The Fraction of Gamma-ray Bursts with an Observed Photospheric Emission Episode}

\correspondingauthor{Zeynep Acuner}
\email{acuner@kth.se}

\author{Zeynep Acuner}
\affil{Department of Physics, KTH Royal Institute of Technology, \\
and The Oskar Klein Centre, SE-106 91 Stockholm, Sweden 
}

\author[0000-0002-9769-8016]{Felix Ryde}
\affil{Department of Physics, KTH Royal Institute of Technology, \\
and The Oskar Klein Centre, SE-106 91 Stockholm, Sweden 
}

\author[0000-0001-8667-0889]{Asaf Pe'er}
\affil{Department of Physics, Bar-Ilan University, Ramat-Gan 52900, Israel}

\author[0000-0002-0041-3783]{Daniel Mortlock}
\affil{Astrophysics Group, Department of Physics, Imperial College London, London SW7 2AZ, UK}
\affil{Department of Mathematics, Imperial College London, London SW7 2AZ, UK}
\affil{The Oskar Klein Centre, Department of Astronomy, Stockholm University, SE-106 91 Stockholm, Sweden
}

\author[0000-0003-4000-8341]{Bj\"orn Ahlgren}
\affil{Department of Physics, KTH Royal Institute of Technology, \\
and The Oskar Klein Centre, SE-106 91 Stockholm, Sweden 
}




\begin{abstract}

There is  no complete description of the emission physics during the prompt phase in gamma-ray bursts. Spectral analyses, however, indicate that many spectra are narrower than what is expected for non-thermal emission models. Here, we reanalyse the sample of 37 bursts in \citet{Yu2019}, by fitting the narrowest time-resolved spectrum in each burst. We perform model comparison between a photospheric and a synchrotron emission model based on Bayesian evidence. We choose to compare the shape of the narrowest expected spectra: emission from the photosphere in a non-dissipative flow and slow-cooled synchrotron emission from a narrow electron distribution. We find 
that the photospheric spectral shape is preferred by $54 \pm 8 \%$ of the spectra (20/37), while $38 \pm 8 \%$ of the spectra (14/37) prefer the synchrotron spectral shape; three spectra are inconclusive. 
We hence conclude that GRB spectra are indeed very narrow and that more than half of the bursts have a photospheric emission episode. 
We also find that a third of all analysed spectra, not only prefer, but are also compatible with a non-dissipative photosphere, confirming previous similar findings.

Furthermore, we notice that the spectra, that prefer the photospheric model, all have a low-energy power-law indices $\alpha \ga -0.5$. 
This means that $\alpha$ is a good estimator of which model is preferred by the data. 

Finally, we argue that the spectra which statistically prefer the synchrotron model, could equally well be caused by subphotospheric dissipation. If that is the case, photospheric emission during the early, prompt phase would be even more dominant.


\end{abstract}



\section{Introduction} \label{sec:intro}


Gamma-ray bursts (GRBs) are bright flashes of gamma-rays emitted from a highly relativistic jet that is ejected during the formation of a black hole in a transient collapse  \citep[see, e.g.,][]{Meszaros2019}.
Within the fireball model of GRBs fluctuations in the jet lead to internal dissipation, which provides energy that can be radiated away.  The observed emission depends on the strength and localisation of such dissipation. However, the smallest radius from which observable emission can originate is from the photosphere, at $r=r_{\rm ph}$, where the jet becomes optically thin. The strength of the photospheric emission depends on the relative amount of energy that is dissipated below the photosphere and on the adiabatic energy losses.
Numerical models of subphotopsheric dissipation show that intense and broad spectra can be produced \citep[e.g., ][]{Rees&Meszaros2005, Peer2005, Giannios2006, Beloborodov2011}. Even without dissipation, angle-dependent outflow properties can add to the broadening of the spectrum in some specific cases, as discussed in \citet{Lundman2013, Ito2013, Meng2019}. 
Photospheric emission naturally produces high radiative efficiency and  a large range of spectral shapes.

Beyond the photosphere, any energy dissipation that accelerates particles and induces magnetic fields would produce synchrotron emission that taps a fraction of the burst energy \citep{Rees1994, Tavani1996}. This is typically assumed to occur at an emission radius of $r_{\rm e} \gg r_{\rm ph}$. Such dissipation could be caused by, e.g., internal shocks, later ejected shells catching up previously emitted shells, or external shocks.  Synchrotron emission is a natural source of radiation in GRBs, both during the prompt and afterglow phases and consistently produces a broad-band spectrum.

Two reasonable emission components are therefore expected during the prompt GRB emission, a photospheric component and a synchrotron component. These can coexist or supersede each other in dominance.  A large variety of possible emission spectra and spectral evolution patterns can therefore be produced depending on varying properties of the central engine as well as on any interactions between the photospheric photons and particles energised above the photosphere \citep[e.g., ][]{Meszaros2002}.

Observationally, there is still not a clear picture of the emission mechanism during the prompt phase of GRBs. This is in part due to the fact that there are ambiguities in the observational determination of the emission components.
Most importantly, the spectral shape of a photosphere, suffering dissipation below the photosphere, 
and a pure synchrotron emission spectrum can resemble each other to such a degree that the available $\gamma$-ray observations cannot clearly distinguish between them. 
In addition, the individual emission components can both be present in the observed spectrum, creating the possibility of misinterpreting the observations,  if not all components are sought for \citep[e.g., ][]{Ryde2005, Axelsson2012, Guiriec2015_New_Model}. 
Finally, in addition to the prompt emission, 
an afterglow component 
is observed to coexist in some bursts \citep[e.g., ][]{Fraija2017, Ajello190114C}. 
 This fact further complicates the interpretation of the observations.

Nevertheless, an important clue to the emission process comes from the shape of the observed spectra, in particular, its width. {The width of the spectrum can, e.g., be defined as the full-width-half-maximum of the $\nu F_{\nu}$-spectrum (power spectrum). In the case of an exponentially cut-off power-law function, the low-energy spectral slope defines the width} \citep[e.g., ][]{Axelsson2015}. As pointed out by \citet{Preece1998}, many spectra have a low-energy spectral slope that is not compatible with slow-cooled synchrotron emission, leading to the so called ``line-of-death'' for synchrotron emission. This was also shown by determining various width measures of the observed spectra \citep{Axelsson2015, Yu2016}. Again a large fraction are too narrow compared to synchrotron emission. 
The estimated fraction differs, though, depending on the method used to determine the width \citep {Burgess2019_Width}. 

In this paper, we revisit the question about the emission process during the prompt phase, studying the earliest prompt emission, and investigate the shape and width of spectrum. In particular, we search for the 
presence of photospheric emission. Our approach will be to statistically compare the observed spectral shape in the {\it Fermi}/GBM energy range to the spectral shapes expected from two specific physical scenarios of synchrotron and photospheric emission. We will use Bayesian model comparison to assess the support for the two physical models.



\section{Sample, Models and Statistical Methods} \label{sec:style}

\subsection{Sample}


Our choice of sample is motivated by our desire to study the intrinsic emission in the early phase of gamma-ray bursts and get unambiguous attributes of the emission process. This task is, however, not straight forward since the GRB emission typically evolves during its duration, with the intensity and spectral peak energy changing \citep[e.g., ][]{Golenetskii1983, Kargatis1994}, but equally important, also changing in the spectral shape and  width \citep[e.g.,][]{Weaton1973, Crider1997}.  These facts have two consequences. First, to observe and study the intrinsic emission, it is vital to study as narrow time interval of the light curve as possible, in order to avoid the observed spectrum to be smeared by spectral evolution. At the same time, though, the signal-to-noise ratio needs to be sufficiently high. Second, a large fraction of the timebins will have spectra that are so broad that, as mentioned in the introduction, they are ambiguous since many emission models can produce them. 

Therefore, the best chance of identifying the intrinsic spectrum is to restrict the investigation to individual pulses in the light curve, which naturally avoids overlapping emission episodes and rapidly varying spectra. Moreover, in order to assess the compatibility of the data with different emission models, the narrowest (intrinsic) spectrum in each burst is of particular interest.

We consequently choose to study the 37 pulses in the catalogue of \citet{Yu2019}.  These pulses were identified as individually connected emission episodes, which can offer sufficiently narrow time-intervals, while maintaining a high statistical significance. This selection  ensures that the spectra represent the intrinsic emission mechanism, by avoiding integration of temporally varying emission. 

\citet{Yu2019} showed that these time-resolved spectra are all well fit with an exponentially cut-off power law function, with a power law index, $\alpha$, which is defined as  
\begin{equation}
N_{\rm E} (E) = K_{\rm CPL}~ \left(\frac{E}{E_{\rm piv}}\right)^{\alpha}~e^{-E/E_{\rm c}},
\label{eq:appCPL}
\end{equation}
where  $N_{\rm E}$ and $K_{\rm CPL}$ is the photon flux and normalisation (cm$^{-2}$ s$^{-1}$ keV$^{-1}$), $E_{\rm c}$ and
$E_{\rm piv} = 100$ keV is the cutoff energy and  pivot energy.
We examine the spectral fits to all of the time intervals in \citet{Yu2019} in order to identify the time intervals which have the largest value of $\alpha$, one from each pulse. 
This selection thus creates a sample of 37 spectra, which all are the narrowest spectra in each individual pulse, and which, as far as possible, represent the intrinsic emission. This sample is presented in Table 1. For the analysis, we used the standard {\it Fermi}/GBM analysis proceedure \citep[e.g. ][]{Goldstein2012, Yu2016}. This includes, for instance, using an energy range of 8 keV  to $\sim$ 850 keV for the NaI detectors, and $\sim 250$ keV to 40 MeV for the BGO detectors, and  using three NaI detectors and one BGO detector, with viewing angles of less than 60 degrees from the location of the burst.
 



\subsection{Emission Models}

Both synchrotron emission and emission from the photosphere could be expected during the prompt emission phase in GRBs \citep[e.g., ][]{Meszaros2019}. These emission processes have in common that they can produce a variety of spectra, with varying widths. However, specific to each model is the theoretical limits to how narrow the spectra are permitted to be.  By comparing the narrowest observed spectra with these two limiting spectral shapes, one can compare the models abilities to explain these narrow spectra.   The primary purpose of this study is therefore to make a model comparison between the  narrowest expected photospheric and synchrotron spectra.

The narrowest synchrotron emission spectrum is expected from 
mono-energetic electrons, which are in a given $B$-field, and  that are in the slow cooling regime. 
However, since any plausible synchrotron-emitting source does not have mono-energetic electrons, we choose to use the emission from a set of electrons with a distribution in energies instead.  
 The theoretical prediction for Fermi acceleration of the electrons provides a power-law index of the electron distribution of either $p=2$ or in the range of $p = 2.2 \mhyphen 2.4$ in the relativistic case \citep{Sironi2015}.  Likewise, the GRB afterglow emission is well described by synchrotron emission from electrons with $p\sim 2$ \citep{WijersGalama1999}.
We know, though, that the theory of particle acceleration is incomplete and we cannot exclude other values at this points.  We therefore choose a larger value of $p=3.8$, which makes the synchrotron spectrum narrower. This value is motivated mainly by the fact that it is the typical value found when slow-cooling synchtrotron spectra are fitted the data \citep[e.g., ][]{Tavani1996, Burgess2019, Ronchi2019}. 

 Moreover, a plausible synchrotron-emitting source is not either expected to have  a single value of $B$ in the emitting region. However, for simplicity, we still retain a single value of $B$, which is an approximation  that avoids an additional broadening of the spectrum and that is typically used in most spectral fits of synchrotron emission.  Finally, we assume an isotropic distribution of 
the electrons' pitch angle, which is the angle between the electron momenta and the magnetic field. In the specific case of an anisotropic pitch angle distribution, the small-pitch-angle synchrotron emission allows for harder spectral slopes below the peak \citep{Epstein1973, Lloyd1999, Medvedev2000, Lloyd2002}. However, it is not clear that such emission is applicable to GRBs \citep{BaringBraby2004, Yang2018}.
In the analysis below we use the emissivity function of synchrotron radiation in a random magnetic field for a power-law distribution of electron energies \citep{NaimaSynch}, implemented in the Python package {\tt Naima} \citep{NaimaPackage}. 
 The emission spectrum used here is shown by the green curve in Figure 1. A power-law with photon index of $-2/3$ is shown by the blue line, which corresponds to  the asymptotic power-law slope for synchrotron emission. This model is referred to as the slow cooling synchrotron model (SCS) below.

Likewise, for the photospheric emission 
the narrowest spectrum is expected in the absence of any significant energy dissipation in the jet that would alter the photon distribution\footnote{Note that if the photosphere occurs while the jet is in the acceleration (photon-dominated) phase 
an even narrower spectrum is expected \citep{Beloborodov2011, Ryde2017}.}. The spectral shape is given by numerical simulations \citep{Beloborodov2010, begue2013monte, Ito2013, Lundman2013} and shown by the red curve in Figure 1 (see also, Fig. 1 in \citet{Ryde2019}). An analytical approximation to the spectrum is given in Equation (1) in \citet{Acuner2019}. This model is called the non-dissipative photosphere model (NDP) below.

\begin{figure}
\gridline{\fig{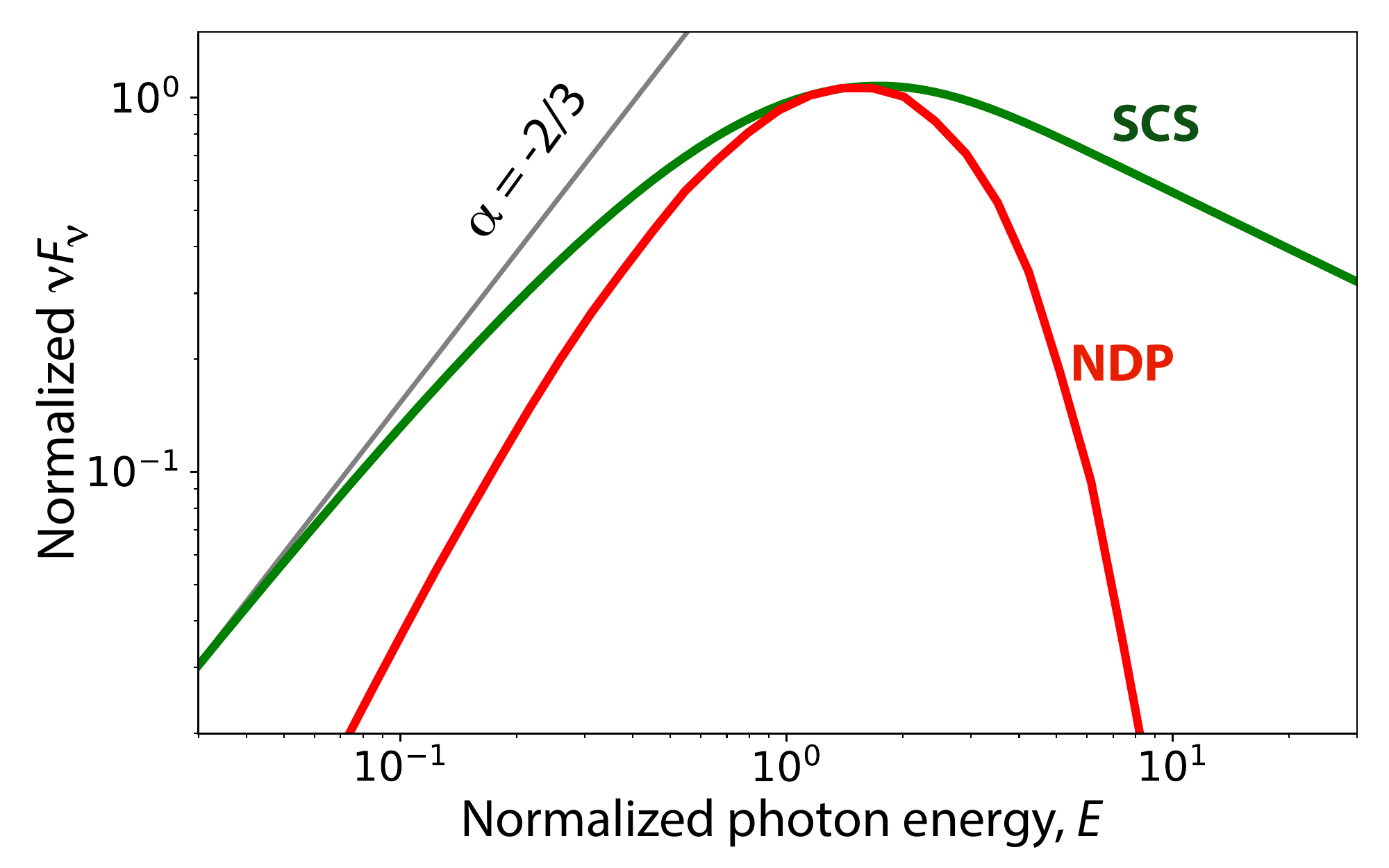}{0.7\textwidth}{}}
\caption{Theoretical spectral shapes that are used to fit to the data, plotted as $\nu F_{\nu}$-spectra. Slow cooled synchrotron emission with electron index $p=3.8$ (green curve) and non-dissipative photosphere emission (red curve). The grey power-law has a power law photon index $\alpha = -2/3$, corresponding to the asymptotic slope of the synchrotron spectrum. \label{fig:spectra}}
\end{figure}

We point out that our primary aim is not to determine the values of the models' parameters, but rather to compare the statistical strength of support for the spectral shapes.
However, the peak energy and the normalisations of the models can be translated into physical parameters \citep{Tavani1996, peer2015}. The ranges of these parameters can serve as additional indications to the validity of the models by comparing them to expected values from theory \citep{Beniamini&Piran2013, Ghisellini2019}. However, to do this, further assumptions are needed about the values of other unknown parameters, such as the bulk Lorentz factor, dissipation efficiencies, and the redshift.

Although the primary goal is to compare the narrowest expected physical emission spectra, we expand the study by including fits with the the empirical cutoff power-law (CPL; Eq. \ref{eq:appCPL}).  CPL is not a physical model, however, it has a flexibility to vary the low-energy power-law slope, $\alpha$, which indicates the shape of the spectrum below the peak. As such the CPL spectrum can characterize expansions of the two emission models, when the strong assumptions that were made for the two  limiting  spectral shapes presented above are relaxed. At the same time, the CPL is a reasonable approximation to some particular physical models such as fast-cooled synchrotron emission \citep[see, e.g., ][]{Burgess2015_alpha} and emission from a photosphere in a photon-dominated flow  \citep{Ryde2017, Acuner2019}. 

\subsection{Bayesian Model Comparison}
\label{sec:ModelComparison}

 We now have a sample of the narrowest spectrum in each burst  which will be fitted to the narrowest allowed theoretical spectra from the two models. The fits to these two models can then be compared statistically. 
In this work, we choose the Bayesian approach for model comparison.

Bayes' theorem  gives the probability of the model ($M$) being true given a set of observed data ($D$) as
\begin{equation}
 P(M \mid D) = \frac{P(D \mid  M) P(M)} { P(D)}, 
\end{equation}
where $P{(M)}$ is the prior probability of the model, $P(D) $ is effectively a normalising factor that ensures the posterior probabilities of the different models sum to unity and $P(D \mid M)$ is the marginal likelihood or Bayesian evidence \citep{jaynesbook}. The marginal likelihood of the $n$th model, denoted $Z_n = P(D | M_n)$ for brevity, is given by
\begin{equation}
Z_{ n} = \int d\theta_{n} \, P(D \mid \theta_{n}, M_{ n}) P(\theta_{n} \mid M_{ n}) {\bf ,}
\label{eq:Z}
\end{equation}
 where $\theta_n$ are the parameters of the $n$’th model, $P(\theta_n | M_n)$ is the prior distribution of the model’s parameters, and $P(D | \theta_n, M_n)$ is the likelihood under the $n$th model given specific parameter values. In other words, the Bayesian evidence $Z$ is the average of the prior-weighted likelihood of a model over the parameter space. The calculation of the evidence thus requires an integration over the whole parameter space and is computationally demanding,  which is one reason why it is not commonly used.  However, recent computational developments make this numerically accessible \citep{skilling, ferozhobson2008}.

 The Bayesian evidence $Z$ is central to the task of model comparison. When Bayesian evidences are obtained for two competing models, the ratio of the respective evidences, ${Z_{2}}/{Z_{1}}$, summarizes the evidence given by the data in favor of model $M_{1}$  or model $M_{2}$ , or equivalently in natural logarithm as the Bayes factor
\begin{equation}
\ln \frac{Z_{2}}{Z_{1}} = \ln\,{{Z_{2}}} - \ln\,{{Z_{1}}}.
\label{eq:K}
\end{equation}
Using this notation, the ratio of the {posterior model probability}  between two models is given as, 
\begin{equation}\label{eq:bayesfactor}
\frac{P(M_{2} \mid D)} {P(M_{1} \mid D)}  = \frac{ P(M_{2})}{P(M_{1})} \cdot  \frac{P(D \mid M_{2})} {P(D \mid M_{1})}.
\end{equation}
When there is no a priori reason to believe that one model is more probable than the other, our prior ratio for the two models becomes unity and the remaining relation at the right hand side equation gives the Bayes' factor \citep[see, e.g., ][]{jaynesbook}.  
Finally, the posterior model probability when the prior ratio is unity is
\begin{equation}
P(M_{2}\mid D) \equiv \frac{P(D \mid M_{2})} {P(D \mid M_{2})+P(D \mid M_{1})} 
= \frac{1} {1 + e^ {-\ln(Z_{2}/Z_{1})}}, 
\label{eq:pmp}
\end{equation}
which gives the probability as a sigmoid function of the Bayes factor (eq. \ref{eq:K}).
As a guide, a difference in log-evidence  greater than $2$ will have a model probability that corresponds to a belief of larger than $90\%$ in the model compared to the other model tested. Between the two models that are compared, we therefore state that a model is preferred if the log-evidence difference $|{\ln (Z_{2}/ Z_{1})}| \gtrsim 2$. The exact value of this threshold is subjective, but similar criteria are typically employed \citep{Jeffreysbook, jeffreysastropaper, kassraftery}.

 In the analysis performed below, we calculate the marginal likelihood or the Bayesian evidence $Z$ (Eq. \ref{eq:Z}) for each spectrum fitted and report the values of $\ln Z_{\rm n}$ for the competing models, their difference in log-evidence; $\ln Z_{2}/Z_{1}$ (Eq. \ref{eq:K}), as well as, the posterior model probability $P$ (Eq. \ref{eq:pmp}).

If this model comparison shows that the non-dissipative photospheric model is statistically preferred over the slow-cooled synchrotron model, then the conclusion can be drawn that the observed spectrum is so narrow that synchrotron model (slow cooled, fast cooled, or marginally fast cooled) is not the best description of the data.
On the other hand, if such a model comparison shows that the slow-cooled synchrotron model is statistically preferred over the non-dissipative photospheric model, then the spectrum is broad enough to be compatible with any synchrotron emission model. However, it cannot be concluded that the emission is not from a photosphere, since the possibility of dissipation below the photosphere can produce a broad spectrum \citep{Peer2005, Beloborodov2011, ahlgren2015confronting}. 
To be able to draw any conclusion in such a case,  a different model comparison must be performed,  between a full synchrotron model and a full subphotospheric dissipation model.  

\subsection{Analysis and Numerical Method}
\label{sec:num}

To sample from the posterior distributions and evaluate evidence integrals, we use {\tt Multinest} which has been shown to successfully and efficiently sample from complicated multimodal posteriors while calculating the evidences in its default run \citep{ferozhobson2008, ferozhobsonbridges}. {\tt Multinest} utilizes the nested sampling (NS) algorithm \citep{skilling}. 
It enables efficient evaluation of the Bayesian evidence where the posterior distribution is given as a by product of the performed run. Therefore, NS requires no additional computations to estimate the marginal likelihood which would be the case for other MC algorithms. Below, we describe in a simplistic way how the algorithm functions.


{\tt Multinest} improves over the standard NS algorithms by utilizing an ellipsoidal rejection sampling scheme which provides better geometrical flexibility for posterior exploration and the ability to identify distinct modes. Due to the nature of the NS method, {\tt Multinest} runs are generally successful in converging on the highest likelihood region. However in some cases, this region might be left outside of the selected elliptical regions during the initial estimations and due to the gradual shrinkage of the initial region the true high likelihood space might be missed, causing misleading posterior distributions. For this reason, we use corner plots to assess the validity of the posteriors. If the posterior distribution does not reflect the expected results in any way, the fits are run multiple times to ensure convergence as well as the live points given to the algorithm are increased so that there is a greater chance to sample the highest likelihood region without eliminating it.

 As the evidence $Z$, given in Eq. (\ref{eq:Z}), is in general computationally expensive to generate, we also calculate the Akaike Information Criterion (AIC) in order to test whether it can be used on this problem.  The AIC is  mathematically simpler measure that could be used as an approximation to evaluate the strength of support for a model. The AIC is defined as
 \begin{equation} 
{\rm AIC} = -2\ln{{\cal L}_{\rm max}}+ 2k,
\label{eq:aic1}
\end{equation}
where $ {\cal L}_{\rm max} = P(D \mid \theta_{\rm max}, M )$ is the maximum likelihood, $\theta_{\rm max}$ is the maximum likelihood parameter values, and $k$ is the number of parameters of the model. Then, the below relation can be written for NDP and SCS models\footnote{Note the number of parameters for NDP and SCS models are the same.},
 \begin{equation} 
{\rm AIC}_{\rm{NDP}} -{\rm AIC}_{\rm{SCS}} = -2\ln{\frac{P(D \mid \theta_{\rm NDP}, M_{\rm NDP} )}{P(D \mid \theta_{\rm SCS}, M_{\rm SCS} )}}.
\label{eq:aic2}
\end{equation} 
where $\theta_{\rm NDP}$ and $\theta_{\rm SCS}$ are the estimated maximum likelihood parameters from the NDP and SCS model fits to the data.

We further utilize posterior predictive checks (PPC) to check for the viability of the model fits. If a model describes the data  well, then the replicated data from the fit should look similar to the observed data that is used to for modelling. Any possible systematic differences in these two distributions hint to the aspects in which the preferred model cannot describe the observed data accurately. 
It should be noted that although PPCs can and do reveal major discrepancies, the related model can still be used for some purposes. In our case, we showcase the model fits with poor PPC results as a further tool for understanding how models can fail to be preferred by data both through evidences and the incapability of reproducing the observational quantities tied to the data \citep{gelman_2004}. We use the two physical models for model comparison without the assumption of any one of them being the standalone processes that might have single-handedly generated our data. Hence, the output of the Bayesian model comparison (performed in \S \ref{sec:ModelPreference}) is not to determine if a certain model is ``correct'' but to see its potential failings and expand the model to encompass these aspects of the data for the future analyses (preformed in \S \ref{sec:expand}). In this regard, PPCs help to pinpoint the exact source of any problematic issues present in the analysis \citep{gelmanvis}. In our analysis, we do not omit the fits with bad PPC results from the presented results. However, this indicates the preferred model is not fully capable of describing the data and hence needs to be modified and/or expanded. This is further discussed in Section \ref{sec:expand}.


\subsection{Selection of Parameter Priors for the Analysis}
\label{sec:priors}

%
%
Priors for each of the two emission models (NDP, SCS) as well as for the empirical CPL model are estimated via the empirical Bayes method. In this procedure, which is a practical step towards more advanced hierarchical modelling, we first derive the approximate priors from the catalogue values and physical arguments. These are used as priors in the {\tt Multinest} fits. The posterior distributions of all the {\tt Multinest} fits for each model then are turned into priors for our actual runs which are the shrunk versions of the initial universal priors assigned. The priors are calibrated this way so that we can assure that we have good sampling and convergence with the given number of live points in a reasonable computational time. By fixing the parameter priors for each model to be the same throughout the sample, we further assure that proper evidences are obtained, which can be affected if different prior intervals are used for different bursts for the same model. We also compare the prior intervals estimated by {\tt Multinest} to those estimated by {\tt emcee} (MCMC) to make sure that we have properly explored the parameter spaces for all models. 

Most priors are distributed as log uniform. The priors for the non-dissipative photosphere spectrum are chosen to be

\begin{equation} \label{eqn:priorsNDP}
\begin{cases}
E_{\rm NDP} \sim \log \mathcal{U}(0.1, 500)\, {\rm keV}\\
K_{\rm NDP}  \sim \log\mathcal{U}(10^{-2}, 2 \times 10^{2})\, {\rm cm}^{-2} {\rm s}^{-1} {\rm keV}^{-1} 
\end{cases}
\end{equation}
where $E_{\rm NDP}$ is  the break energy and $K_{\rm NDP}$ is its normalisation, and $\mathcal{U}(x_{\rm min}, x_{\rm max})$ is a uniform and  $\log\mathcal{U}(x_{\rm min}, x_{\rm max})$ is a log-uniform distribution of the variable $x$, between $x_{\rm min}$  and $x_{\rm max}$.   For the slow-cooled synchrotron spectrum, we use the correpsonding priors 
\begin{equation} \label{eqn:priorsSCS}
\begin{cases}
E_{\rm SCS} \sim \log \mathcal{U}(1, 2000)\, {\rm keV}\\
K_{\rm SCS}  \sim \log\mathcal{U}(10^{-2}, 10^{2})\, {\rm cm}^{-2} {\rm s}^{-1} {\rm keV}^{-1} 
\end{cases}
\end{equation}
Finally, the empirical cutoff powerlaw model, which is defined in Equation (\ref{eq:appCPL}), has the following priors: 
\begin{equation} \label{eqn:priorsCPL}
\begin{cases}
E_{\rm CPL} \sim \log \mathcal{U}(10, 1500)\, {\rm keV}\\
K_{\rm CPL}  \sim \log\mathcal{U}(10^{-2}, 5 \times 10^{2})\, {\rm cm}^{-2} {\rm s}^{-1} {\rm keV}^{-1}\\
\alpha \sim \mathcal{U}(-2, 1.5)
\end{cases}
\end{equation}
We note that the prior ranges differ somewhat, simply due to the differences in spectral shape, and their definitions. For instance, the parameters $ E_{\rm NDP},E_{\rm SCS} $ and $E_{\rm CPL}$ are the break energies of the respective functions and hence are not identical with each other nor with the peak energy, $E_{\rm pk}$. 

\subsection{Sensitivity to Priors}

  The AIC value (Eq. \ref{eq:aic1}) evaluates the best fit likelihood and is, unlike the evidence (Eq. \ref{eq:Z}), insensitive to the parameter priors assuming the true likelihood maximum is within the parameter range encompassed by the priors. Therefore, we will compare these two measures to see if they differ, in order to assess whether it is reasonable to approximate the full evidence calculation with the AIC. The AIC values are shown in Table  \ref{tab:appendix} in the Appendix.

In Figure~\ref{fig:EvAIC}, we compare the corresponding differences in log-evidence and AIC for the fits to the NDP and the SCS models.  A positive difference in log-evidence indicates that the NDP model is favoured, while the opposite is true for the AIC. The figure shows that the AIC differences are strongly correlated with the difference in log-evidences. It is therefore evident that, in this case, the AIC measure captures 
everything important in the fit.  The correlation also indicates that our results are not critically sensitive to the  prior choices, since the AIC is insensitive to the parameter priors. We conclude that, at least for the priors we have adopted and with the data and model set described, the AIC difference is a reasonable approximation to the difference in log-evidences. 

The robustness of the priors have also been tested by assigning  different sized prior ranges to the fits and checking the fits and evidence values. We use the smallest possible prior ranges that include the greatest likelihood region in order to facilitate the numerical estimation of the Bayesian evidence.
We also make an effort of approximately matching the prior ranges across parameters for the three different models used (see \S \ref{sec:priors}).

\section{Results}


In order to compare the two models (NDP and SCS), we calculate the Bayesian evidence ($Z$, or marginal likelihood, Eq. \ref{eq:Z}) for each of the models, assuming the prior distributions specified in \S~\ref{sec:priors}.
In Table 1, we show the results of the fits to the 37 spectra in our sample. The GRB number ID 
are followed by the log-evidence values ($\ln Z$) of the two models, as well as their differences, $\ln (Z_{\rm NDP}/ Z_{\rm SCS})$ (Eq. \ref{eq:K}), and corresponding posterior model probability, $P$ (Eq. \ref{eq:pmp}). For $\ln (Z_{\rm NDP}/ Z_{\rm SCS}) > 0$, $P$ is the probability for the NDP model to be the better model, while for $\ln (Z_{\rm NDP}/ Z_{\rm SCS}) < 0$, $P$ is probability for the SCS model to be the better model. Consequently, $P$ is always larger than  $0.5$.
In Table \ref{tab:appendix}, in the Appendix, we include more information of the analysed spectra. 



\begin{table}[ht]
\centering
\begin{tabular}{ccccccc}
  \hline
  GRB ID & $\alpha $ & $\ln Z_{NDP}$ & $\ln Z_{SCS}$ & $\ln (Z_{\rm NDP}/ Z_{\rm SCS})$  & $P$ \\ 
  \hline
  081009140 & -0.47$\pm$0.18& -627.88 & -730.35 & 102.48 & 0.99   \\ 
081125496 & -0.44$\pm$0.17& -1406.15 & -1409.19 & 3.04 & 0.75 & \\ 
081224887 & -0.09$\pm$0.08 & -1526.86 & -1655.56 & 128.70 & 0.99   \\ 
090530760 & -0.13$\pm$0.18 & -2033.84 & -2065.46 & 31.61 & 0.97  \\ 
090620400 & 0.10$\pm$0.13 & -1571.63 & -1667.18 & 95.55 & 0.99  \\ 
090626189 & -0.61$\pm$0.11 & -260.62 & -254.85 & -5.77 &   0.85 \\ 
  090719063 & 0.18$\pm$0.13 & -726.33 & -838.66 & 112.33 & 0.99   \\ 

   090804940 & -0.09$\pm$0.28 & -419.45 & -439.52 & 20.07 & 0.95  \\ 
 090820027 & -0.45$\pm$0.09 & -122.68 & -140.43 & 17.75 & 0.95   \\ 
100122616 & -1.40$\pm$0.12 & -1293.80 & -1236.44 & -57.37 &  0.98 \\ 
  100528075 & -0.95$\pm$0.09 & -1066.25 & -1020.59 & -45.66 &   0.98 \\ 
  100612726 & -0.12$\pm$0.15 & -199.77 & -234.78 & 35.01 & 0.97 & \\ 
   100707032 & 0.27$\pm$0.03& -1892.17 & -2905.26 & 1013.10 & 1.00   \\ 

 101126198 & -1.15$\pm$0.08 & -2289.30 & -2224.14 & -65.16 &   0.98 \\ 
  110721200 & -0.94$\pm$0.04 & -1449.86 & -1158.11 & -291.74 &   1.00 \\ 

 110817191 & -0.49$\pm$0.14 & -869.74 & -870.45 & 0.71 & 0.42   \\ 
   110920546 & 0.28$\pm$0.19 & -3521.46 & -3610.30 & 88.84 & 0.99   \\ 

111017657 & -0.76$\pm$0.07 & -1238.29 & -1170.10 & -68.19 &  0.99 \\ 
 120919309 & -0.65$\pm$0.06 & -966.25 & -925.93 & -40.32 &   0.98 \\ 

130305486 & -0.38$\pm$0.08 & -1044.69 & -1071.74 & 27.05 & 0.96   \\ 
  130612456 & -0.78$\pm$0.07 & -1122.34 & -1115.38 & -6.97 &  0.87 \\ 

 130614997 & -1.21$\pm$0.12 & -1040.20 & -998.83 & -41.38 &   0.98 \\ 

 130815660 & -0.66$\pm$0.13 & -612.87 & -613.20 & 0.33 & 0.25  \\ 
   140508128 & -0.63$\pm$0.06 & -500.92 & -440.88 & -60.04 &   0.98 \\ 

   141028455 & -0.65$\pm$0.09 & -1370.18 & -1346.76 & -23.42 &  0.96 \\ 

141205763 & -0.69$\pm$0.10 & -1454.90 & -1453.80 & -1.10 &   0.52 \\ 
  150213001 & -0.97$\pm$0.05 & -820.94 & -698.59 & -122.35 &   0.99 \\ 
  150306993 & -0.06$\pm$0.18  & -978.49 & -1011.36 & 32.86 & 0.97  \\ 


150314205 & -0.23$\pm$0.08 & -237.69 & -319.26 & 81.57 & 0.99   \\ 
  150510139 & -0.53$\pm$0.06 & -70.74 & -67.43 & -3.31 &   0.77 \\ 
 150902733 & -0.30$\pm$0.04 & -813.65 & -1026.51 & 212.86 & 1.00   \\

 151021791 & -0.18$\pm$0.13 & -497.58 & -537.32 & 39.74 & 0.98   \\ 
  160215773 & -0.86$\pm$0.04 & -2397.69 & -2258.47 & -139.22 &   0.99 \\

160530667 & -0.51$\pm$0.02 & -1723.88 & -1792.52 & 68.64 & 0.99   \\ 
160910722 & -0.12$\pm$0.15 & -293.08 & -333.11 & 40.04 & 0.98  \\ 

 161004964 & -0.28$\pm$0.15 & -1573.30 & -1593.07 & 19.77 & 0.95   \\ 

  170114917 & -0.45$\pm$0.08 & -886.54 & -900.68 & 14.13 & 0.93  \\ 

   \hline
\end{tabular}
\caption{CPL low-energy index ($\alpha$), log-evidences for NDP and SCS, the Bayes factor $\ln (Z_{\rm NDP}/Z_{\rm SCS})$, as well as the  posterior  model probability $P$ from eq. (\ref{eq:pmp}) are presented. For $\ln (Z_{\rm NDP}/ Z_{\rm SCS}) > 0$, $P$ is the posterior model probability for the NDP model, while for $\ln (Z_{\rm NDP}/ Z_{\rm SCS}) < 0$, $P$ is the posterior model probability for the SCS model. Consequently, $P$ is always larger than  $0.5$.}\label{tab:table1}
\end{table}

\begin{figure}
\gridline{\fig{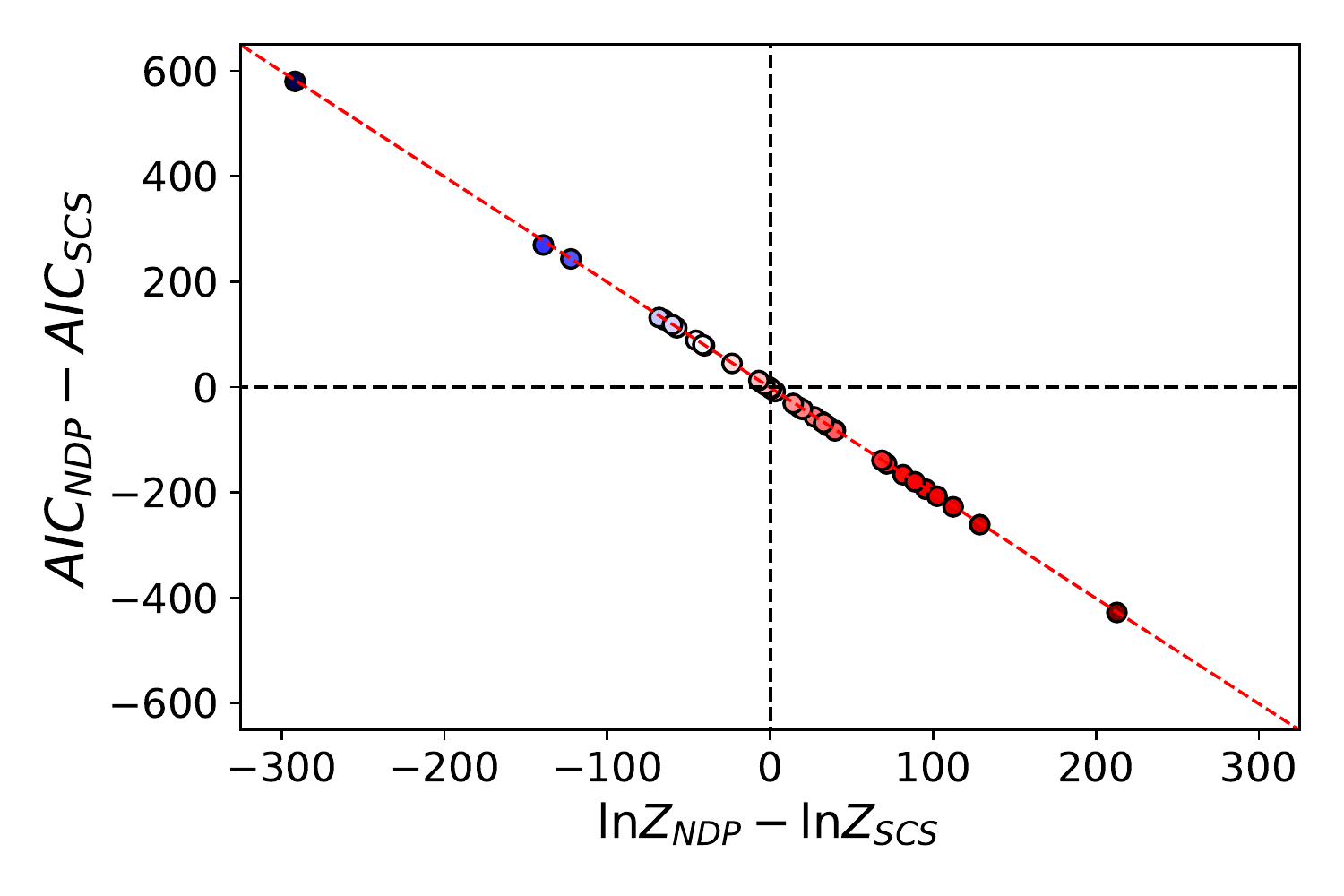}{0.7\textwidth}{}}
\caption{Model comparisons between two physical models, the NDP model and the SCS model. The $y$-axis gives the difference in Akaike Information Criterion (AIC) of the fits as described in Eq. (\ref{eq:aic2}) and the $x$-axis gives the difference in the corresponding Bayesian evidence as $\ln Z_{\rm NDP} - \ln Z_{\rm SCS}$ [see Eq. (\ref{eq:K})]. The range of colors corresponds the range of the values of $\ln (Z_{\rm NDP}/Z_{\rm SCS})$. GRB 100707A has a very large $\ln ( Z_{\rm NDP}/Z_{\rm SCS}) = 1013$ and lies on the same line, however, it is left out of the plot for clarity.
 \label{fig:EvAIC}}
\end{figure}


\subsection{Model Preference}
\label{sec:ModelPreference}

We will now discuss the differences in log-evidence, $\ln (Z_{\rm NDP}/ Z_{\rm SCS})$, shown in Table \ref{tab:table1}.  As mentioned above, we choose to use the notion that a model is preferred if $|\ln (Z_{\rm NDP}/ Z_{\rm SCS})| \gtrsim 2$. There are only three spectra that 
have $|\ln (Z_{\rm NDP}/ Z_{\rm SCS})| \lesssim 2$ and that therefore are excluded in the discussion\footnote{GRBs 110817, 130815, and 141205}, 
since they are indecisive.
Table \ref{tab:table1} shows consequently that for {20/37} (54\%) of the analysed spectra, the photosphere model is preferred, since $\ln (Z_{\rm NDP}/ Z_{\rm SCS}) > 2$. Similarly, for {14/37} (38\%) of the spectra, the synchrotron model is preferred, since  $\ln (Z_{\rm NDP}/ Z_{\rm SCS})< -2$. 


We note that for most of the bursts the difference is log-evidence, $\ln (Z_{\rm NDP}/ Z_{\rm SCS})$, is large. This means that we can claim that one of the models is preferred over the other with great confidence, but not that either model is necessarily a good fit to the data.  Table \ref{tab:table1} shows that the bursts with the two largest difference in evidences  are GRB100707, {overwhelmingly} in favour of the photosphere model, and GRB110721A, {overwhelmingly} in favour of the synchrotron model. Apart from these burst, there are four bursts with $\ln (Z_{\rm NDP}/ Z_{\rm SCS}) > 100$ and one with  $\ln (Z_{\rm NDP}/ Z_{\rm SCS}) < -100$ (see Table \ref{tab:table1}).  In the Appendix, we present some of the analysis products, such as the PPC plots and the spectral fits, for two  bursts (GRB150314 and GRB150902) to illustrate the analysis performed and what good and bad fits typically look like. 

We proceed to  investigate whether the model preference depends on the $\alpha$-value found from fits to the traditionally used empirical model, the cut-off power law function \citep[CPL, ][]{Band1993}. This spectrum is characterized by the low-energy power law index, $\alpha$ (see Eq. \ref{eq:appCPL}). Therefore, in our analysis we have also refit the spectra to the CPL function.  For our spectra, such fits are already presented  in \citet{Yu2019} and \citet{Ryde2019},  who used {\tt MCMC}  for posterior sampling. Here, instead, we use {\tt Multinest} as a sampler (\S \ref{sec:num}) in order to be able to be able to calculate the CPL fit evidences (used in \S \ref{sec:expand}). We find similar fit results to within errors for all the analysed spectra. 

\begin{figure}
\gridline{\fig{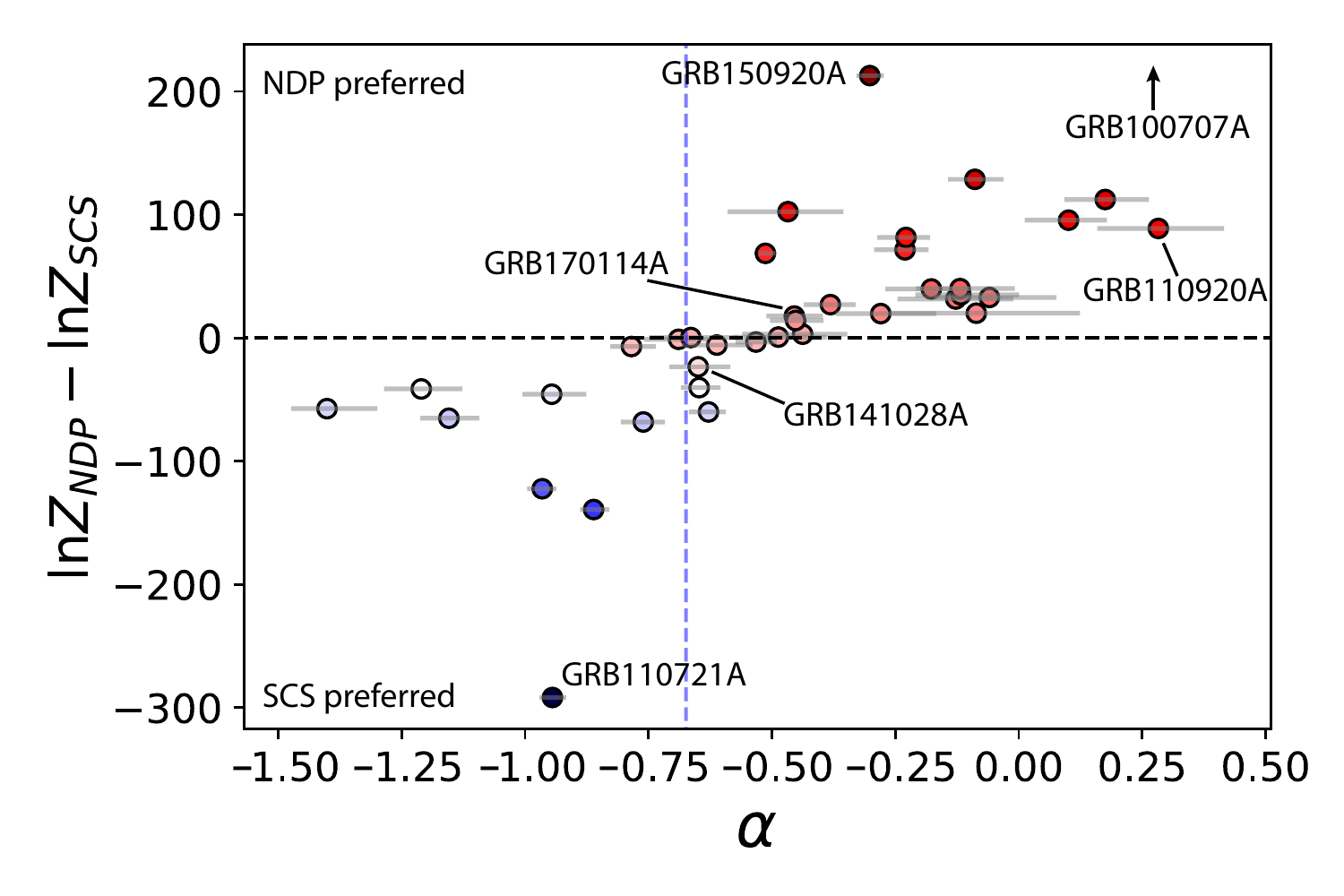}{0.7\textwidth}{}}
\caption{Difference in log-evidence, between the fits to the NDP model and the fits to the SCS model, versus the low-energy power-law index, $\alpha$, of the cutoff powerlaw function. 
Positive values of the log-evidence difference (corresponding to $\ln (Z_{\rm NDP}/ Z_{\rm SCS})$ in Eq. [\ref{eq:K}]) indicate a preference for the NDP model. The vertical, dashed, line indicates $\alpha = -2/3$. 
The burst names of a few particular bursts, that are discussed in the text, are indicated at the individual data points. GRB 100707A has a very large $\ln ( Z_{\rm NDP}/Z_{\rm SCS}) = 1013$ and its position is therefore only indicated by an arrow.
\label{fig:Evi_vs_alpha}}
\end{figure}

In Figure \ref{fig:Evi_vs_alpha}, we plot the difference in log-evidence between NDP and SCS versus the corresponding value of $\alpha$. In the plot the line of $\alpha =-2/3$ is indicated, which corresponds to the asymptotic power-law photon index for the SCS model. 
Figure \ref{fig:Evi_vs_alpha} shows that there is a strong correlation between $\ln (Z_{\rm NDP}/ Z_{\rm SCS})$ and $\alpha$.  When the NDP is the preferred model, then the $\alpha$-value typically is  $\ga -0.5$, while for the cases where $\alpha$ is $\la -0.6$, then the SCS is systematically the preferred model, and finally the three indecisive cases lie between $ -0.7 \la \alpha \la -0.5$. This means that the $\alpha$-parameter can be used as an approximate determination of the model preference of the data. Eluding to the `line-of-death' of synchrotron emission \citep{Preece1998}, we can thus define a `line of non-compatability' of synchrotron emission,  lying at $\alpha = -0.5$, instead.


\subsection{Expanding Beyond the Two Models}
\label{sec:expand}

In the analysis above we have compared the strength of support  for two specific models, NDP and SCS and determined which model is the preferred one. However, the actual property of the burst emission could be different from any of the tested models.  Natural expansions  include subphotospheric dissipation for the photosphere model \citep[see, e.g.][]{Peer2005, Beloborodov2011, Ryde2011} and include different cooling regimes and evolutions in the case of synchrotron emission \citep[see, e.g., ][]{Tavani1996, Lloyd2000, Burgess2019}. A way to investigate this, is to compare the fits of the physical models to fits to the empirical cutoff powerlaw (CPL) model, eq. (\ref{eq:appCPL}). The flexibility of this function is given by the power-law slope, $\alpha$, which can capture all of the model expansions mentioned above.  The log-evidence values found from fitting our spectra with the CPL model are given in Table \ref{tab:appendix}.

\begin{figure}
\gridline{\fig{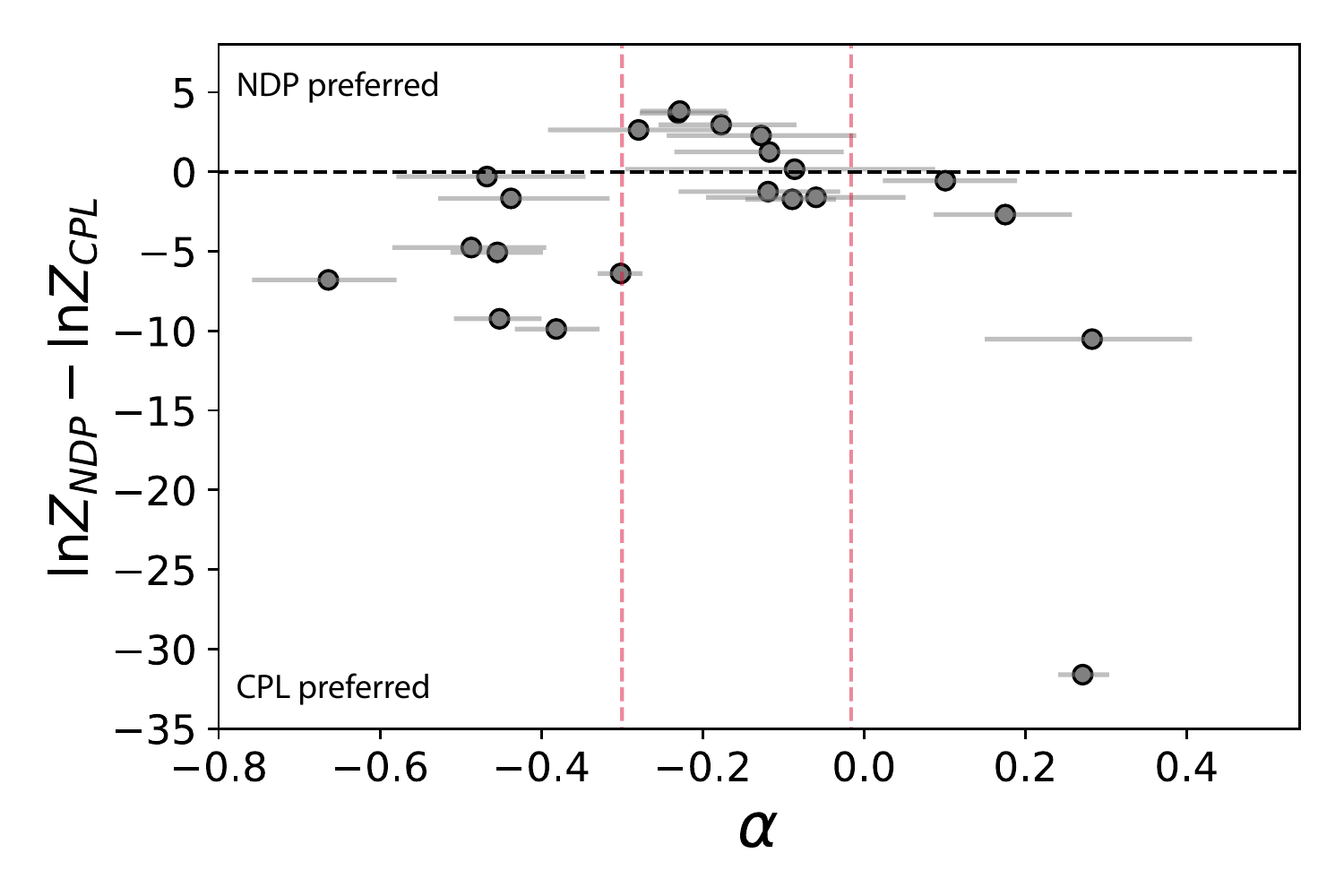}{0.7\textwidth}{}}
\caption{Difference in log-evidence between the fits to the NDP model and the fits to the CPL model versus the low-energy power-law index, $\alpha$, of the CPL model. This figure only includes the 22 cases in Fig. \ref{fig:Evi_vs_alpha}, for which $\ln (Z_{\rm NDP}/ Z_{\rm SCS}) > 0$.
Positive values of the log-evidence difference indicate a preference for the NDP model. The red, dashed lines indicates the $\alpha$-interval in which CPL fits to NDP generated spectra would lie \citep{Acuner2019}.
 \label{fig:CPLNDP}}
\end{figure}

We now focus on the 22 burst spectra which have a preference for the NDP over SCS.
The bursts are shown in Figure \ref{fig:Evi_vs_alpha} above the dashed line since they have $\ln (Z_{\rm NDP}/ Z_{\rm SCS}) >0$. In Figure \ref{fig:CPLNDP} we plot the difference in log-evidence between the NDP model and the CPL model as a function of the CPL parameter $\alpha$. 
The figure shows that there are six bursts with positive difference in $\ln Z_{\rm NDP} - \ln Z_{\rm CPL}$ (Table \ref{tab:appendix}). This means that the physical NDP model is preferred over the empirical CPL spectrum, even though the latter has more flexibility by having one more free fit parameter. Most of  these cases occur in an interval around $\alpha \simeq -0.2$. 
%
%
%
Moreover, Figure \ref{fig:CPLNDP} shows that for  $\alpha \lesssim -0.3$ the CPL is the preferred model for most cases. The same is for the  spectra with $\alpha \gtrsim 0$. We interpret this as the spectra are not fully correctly represented by the NDP, in particular, the low-energy slope is not correct. 

A natural explanation for this is that the $\alpha \la -0.3$-cases are better produced by subphotospheric dissipation, which naturally produces a soft low-energy shoulder of the spectrum  \citep[see, e.g.][]{Peer2005, Ahlgren2019}. Of the spectra that have  $\alpha \leq -0.3$ there are five cases for which the two requirements hold, namely $\ln (Z_{\rm NDP}/ Z_{\rm SCS}) > 2$ as well as $\ln (Z_{\rm NDP}/Z_{\rm CPL}) < -2$. These five spectra can thus be claimed to be  due to subphotospheric dissipation\footnote{GRBs 090820, 130305, 150902, 160530, and 170114.}.


Likewise, it can also be argued that the cases with $\alpha \gtrsim 0$ are better fit by a spectrum that is even narrower than the NDP \citep[see, e.g., ][]{Ryde2017, Acuner2019}. The only possibility for this to occur within the fireball model is if the photosphere takes place in the acceleration phase, before the flow saturates \citep{Beloborodov2011, Ryde2017, Chhotray2018}. 
Since dissipation is not expected below the saturation radius, such spectra are naturally non-dissipative. Therefore, together with the spectra that prefer NDP (with positive difference in $\ln Z_{\rm NDP} - \ln Z_{\rm CPL}$) we therefore can argue that all the 13 bursts with $\alpha > -0.3$ are consistent with being non-dissipative. 

We note that the fraction of non-dissipative spectra (13/37 = 35\%) that we find here is similar, but somewhat larger, than the fraction of spectra consistent with non-dissipative photospheres found by \citet{Acuner2019}.  They used a different approach by analysing synthetic spectral data, which were produced by simulating data from the theoretical NDP spectrum and then convolved through the GBM response. 
\citet{Acuner2019} showed that when such synthetic data, having a range of spectral peak energies, are fitted with a CPL function it results in  $-0.5 \la \alpha \la 0.1$.
 By comparing this range with  the full catalogue values of $\alpha$, \citet{Acuner2019} found that 1/4 of all are consistent with NDP. The FWHM of this distribution (blue curve in Fig. 5 in \citet{Acuner2019}) is shown by the dashed, red lines in Figure \ref{fig:CPLNDP}.  Most spectra in this range do have positive difference in log-evidence in Figure \ref{fig:CPLNDP}.
 One obvious difference in our work, which could explain the small difference in fractions, is that we only analysed one spectrum per bursts while \citet{Acuner2019} compared to the $\alpha$-values in the full catalogue of \citet{Yu2016}. However, it is worth pointing out that these two different methods, one based on comparing simulated spectral shapes to the catalogue shapes and the other based on model comparison,  give  similar results. This consolidated the conclusion that a non-negligible fraction of GRB spectra are consistent with photospheric emission in a non-dissipative flow. 


In summary, our statistical analysis and consideration concludes that, not only 20/37  ($54\%$) of the analysed spectra are  compatible with a  photospheric origin, but also that  13/37 ($35\%$  ; Fig. \ref{fig:CPLNDP}) are actually compatible with photospheric emission in a non-dissipative jet. 




\section{Discussion}

We are addressing the fundamental mechanism of  the emission during the prompt phase in GRBs. In order to do this we study the  narrowest spectra in 37, well-detected, GBM  bursts. We discriminate between the radiation processes by comparing the spectral shapes of a non-dissipative photosphere (NDP) spectrum and a slow-cooled synchrotron (SCS) spectrum. 

In the choice between NDP and SCS,  we find that 54 $\pm$ 8 \% of the  spectra prefer the photospheric model while 38 $\pm$ 8 \% prefer the sychrotron model\footnote{Errors on the percentages are standard errors with sample size $N=37$.}.  It is important to bear in mind that the statistical answer we arrive at is the spectra's preference between these two models only. This means that we did not, necessarily, arrive at the best model for the spectrum, which could be a different model.  Indeed, we showed, in \S \ref{sec:expand}, that spectra with $-0.6 \la \alpha \la -0.3$ still do prefer the NDP model over the SCS model, but there must be some additional broadening mechanism to explain the spectra fully. Therefore, we argued that a subphotospheric dissipation model is an even better description of the data, compared to NDP and SCS, for these cases.

\subsection{Subphotospheric Dissipation when $\alpha \la -0.6$}

The fact that a broadening mechanism is required for some of the spectra with $\alpha \ga -0.6$ leads to the natural consequence that a fraction of the rest of the sample (spectra preferring the SCS over NDP) also could be from the photosphere. The reason is that  if  a photospheric broadening mechanism is needed for some bursts, there is no reason to believe that it is limited to $\alpha \ga -0.6$. On the contrary, values down to $\alpha \sim -1.1$ are theoretically expected \citep{TompsonGill2014, Vurm2016}.   There are only three spectra\footnote{GRBs 100122, 101126, 130614.} in Table \ref{tab:table1} that, given the uncertainties, are consistent with having $\alpha < -1.1$.  Since these three spectra are not naturally formed by subphotospheric dissipation, it is difficult to argue for anything but prompt synchrotron emission for them.
This deliberation points to a natural possibility that a significant majority of the analysed spectra are indeed photospheric and only a small fraction  are due to  synchrotron spectra: $35\%$ (13/37) are from non-dissipative photospheres, 57\% (21/37) are from dissipative photospheres, and  8\%  (3/37) are from synchrotron emission.


\subsection{Similar Spectra: Synchrotron and Photosphere} 
\label{sec:SynPhot}

The deliberation in the previous section argued that many of the spectra preferring the SCS over NDP actually could be photospheric. Indeed, the  spectral shape of synchrotron emission and subphotopsheric dissipation spectrum can be similar in shape. To illustrate this point, we will now create synthetic spectra from a specific scenario within the subphotospheric dissipation model and then fit them  with a synchrotron model.

Synchrotron spectra are, by nature, very broad, emitting over a large energy range. However, spectra formed by subphotospheric dissipation can also be broad. They are, though, inherently marked by a low-energy break, relating to the seed photons for the unsaturated Comptonization up to the peak energy. The position of the low-energy break depends on the exact prescription of dissipation, but can be several decades in energy below the main peak \citep{peer2015, Vurm2016, TompsonGill2014, BhattacharyaKumar2019, Ahlgren2019}.  Further, such a spectrum will be marked by a high-energy cutoff at around $\sim 50-100$ MeV due to intrinsic pair opacity \citep{Peer2005, Beloborodov2011}. In any case, over the limited energy range of the GBM detector the subphotospheric spectrum can resemble a Band spectrum, or even a synchrotron spectrum.

\begin{figure}
\gridline{\fig{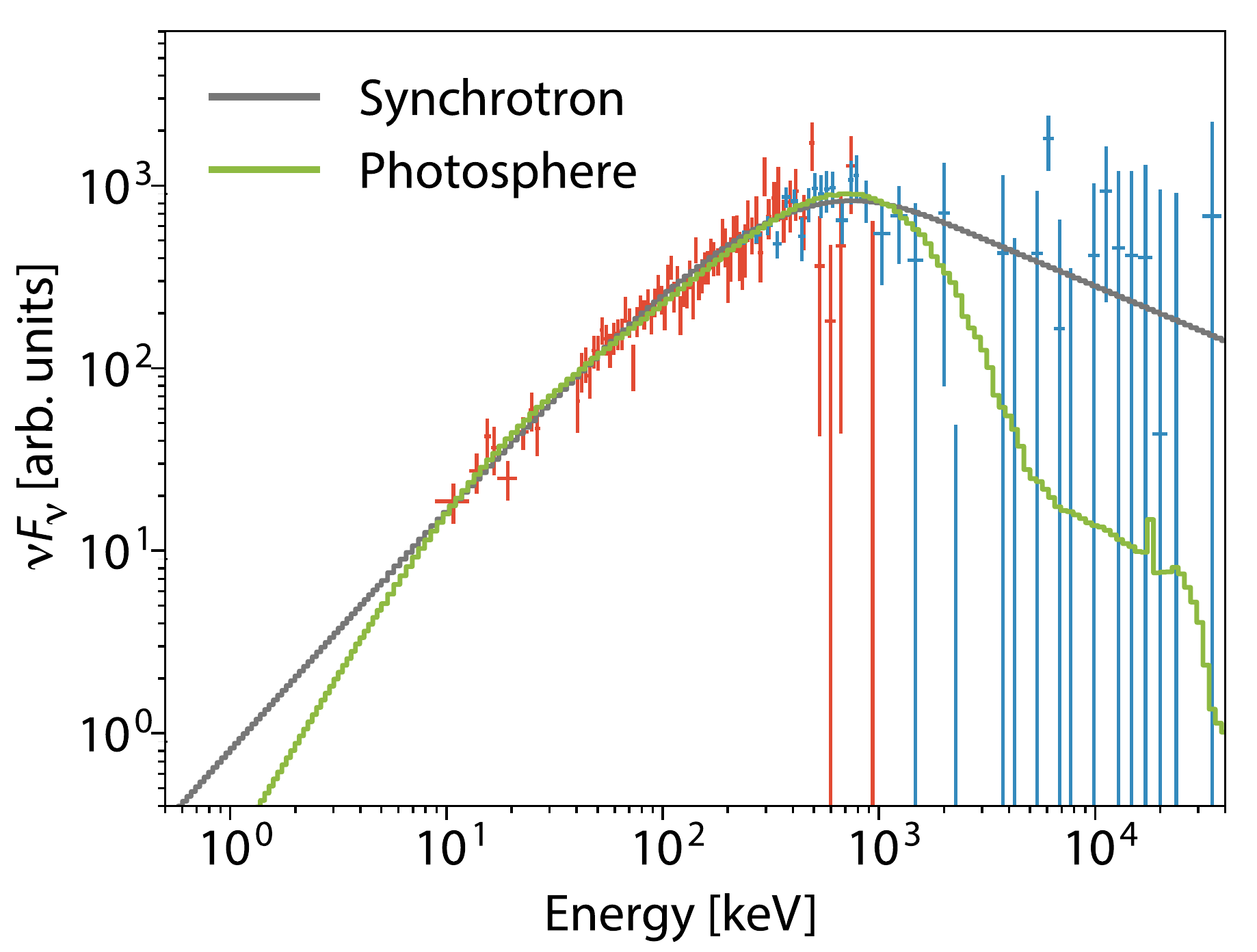}{0.7\textwidth}{}}
\caption{Synthetic data from a subphotospheric dissipation model (green curve) for photospheric emission. The  grey curve shows the synchrotron spectral fit in the GBM energy interval, which gives a good fit.
The fit is performed in {\scriptsize XSPEC} excludes data in the region of $30 - 40$~keV around the $K$-edge. Red data points are synthetic data from the NaI detector and blue data points are from the BGO detector.
\label{fig:Bjorn}}
\end{figure}

Our synthetic spectra  are produced using the model of \citet{Peer2005}. One such spectrum is shown by the (synthetic) data points in Figure \ref{fig:Bjorn}. 
The data are generated from a model (green curve), with a small amount of subphotospheric dissipation in a high luminosity jet. The model parameters are typical for other observed bursts that are fitted by the model \citep{ahlgren2015confronting, Ahlgren2019}.
The data are generated 
using {\scriptsize XSPEC} \citep{XSPEC1999} and have a signal-to-noise ratio of $\sim 20$. 
As shown by the figure, the fit to the synchrotron model (grey curve) gives a good representation of the data. The AIC value for the synchrotron fit is 273.2, while the AIC value for a fit to the photospheric model itself gives 273.4, which are not significantly different. The data at high energies are in this case not enough to reliably discriminate between the two models, as is often the case. However, data at energies $\la 10$~keV could in this instance help distinguish between the models.  By inspection of this fit alone it is not possible to tell that the data have been generated from a scenario of subphotospheric dissipation.


\subsection{Compatible Model versus Preferred Model}

Turning back to the analysed spectra, which were the hardest spectra in each burst, we found that $38\%$ of the spectra strongly prefer a synchrotron spectrum over the NDP. Indeed, 
many works have identified GRB spectra that are compatible with synchrotron emission \citep[e.g., ][]{Tavani1996, Lloyd2000, Burgess2011, Oganesyan2018, Burgess2019}.  

Again, from a statistical perspective, all these previous findings do not prove the synchrotron model, but they show that the data is compatible with a certain spectral model.  
The spectral fit in Figure \ref{fig:Bjorn} illustrates this inherent ambiguity of the GBM data to clearly distinguish between the two physical models. 
A further example is GRB170114A which  provides a good fit to a slow-cooling synchrotron  model \citep{Burgess2019_170114}. In our analysis above, 
we found a clear preference for NDP for this burst (Table 1; $\ln (Z_{\rm NDP}/ Z_{\rm SCS})$ = 14). This means that there is a physical model that is statistically preferred by the data over the SCS model. The conclusion of this fact is that the emission in GRB170114A is not synchrotron but it is from the photosphere.
On the other hand, including the cut-off powerlaw function (CPL) in the model comparison (see \S \ref{sec:expand}), the CPL becomes the preferred model for GRB170114A. This is mainly due to the fact that $\alpha \sim -0.5$, which is very soft for photospheric emission. The conclusion we draw from this  analysis is that the best physical model for (largest $\alpha$-timebin in) GRB170114A is a dissipative photospheric emission.

\subsection{Afterglow Overlapping and Superseding the Prompt Emission}

The spectra we analysed in this paper are from pulses in the light curve, which are mainly in the vicinity of the GBM trigger, within $\sim 10$ s \citep[see Table 1 and ][]{Yu2019}. Only eight of the spectra occurred more than 10 s after the trigger. {Since the afterglow onset timescale is} on the order of a few tens of seconds, 
 the afterglow component could be expected in observations much beyond this time scale.

This point is further underlined by the fact that many burst are observed to have a second emission component which is consistent with being synchrotron emission and that coexists with the prompt emission.  In particular,  observations of high-energy, LAT photons indicate that they are typically delayed relative to the GBM trigger, but emerge while the GBM emission is still active \citep[e.g., ][]{ 090510, Giuliani2014, Ajello2019}. 
{ These components have been associated to the onset of the afterglow emission from an external blastwave, involving the reverse shock and forward shocks \citep[e.g., ][]{Kumar2009, Ghirlanda2010, Panaitescu2017}. Alternatively, such extra components have been successfully explained to  be due to inverse Compton cooling of the heated $e^\pm$ enriched plasma, created behind the forward shock in the blastwave, due to the exposure of the prompt keV-MeV emission \citep{Beloborodov2014, Vurm2014} and that involves two phases, an inverse Compton phase and a synchrotron-self-Compton phase.}
Recently, a very clear example of such emission is given in GRB190114C \citep{Ravasio2019, Ajello190114C}. Here a synchrotron component decidedly emerges during the prompt phase and  this component is unmistakably connected to the  synchrotron component of the early afterglow emission of the blastwave \citep{Fraija2019, DerishevPiran2019}.  In the case of GRB190114C  the synchrotron component is initially observed to fluctuate until it finally settles into the temporal power-law of the long term afterglow. The afterglow component therefore coexists with the prompt emission in many bursts. 

In other cases, the observed photospheric emission is superseded by a synchrotron component. 
Examples are GRB150330A and GRB160625B, which have an initial, short photospheric episode, while their main emission episode occurs a few hundred seconds later, which are dominated by synchrotron emission \citep{ZhangBB2018, Li2019BBPointing}. Other similar cases have the components separated by a few 10s of seconds instead \citep[GRB140329B and GRB160325A; ][]{Li2019BBPointing, Sharma2020}. 
This type of bursts illustrate the fact that if the initial thermal pulse had been missed for some reason, they would have been detected with a purely synchrotron-like prompt emission. 



It was, in fact, early argued \citep{ReesMeszaros1992} that prompt emission might be absent and that all observed emission could be synchrotron emission from the external shock, i.e., the afterglow \citep[see further discussion in ][]{Panaitescu1998}. An example of such a case is GRB141028A, where all the observed emission, both the "prompt" phase 
as well as the afterglow phases, are interpreted to be from one and the same process, which is consistent with an external shock \citep{Burgess2016}. {In the analysis above, this particular burst has  $\ln (Z_{\rm NDP}/ Z_{\rm SCS}) = -23$ (Table 1), indicating a strong preference for the synchrotron model, thus supporting such an interpretation.} A similar suggestion
was made for the interpretation of GRB110721A \citep{Iyyani2016}, which had $\ln (Z_{\rm NDP}/ Z_{\rm SCS}) < - 200$ above.

The possibility of observing only  the external shock emission is in line with the fact that when synchrotron spectra are found to be compatible with the prompt emission, a general finding from the fits is that the emission radius, the bulk Lorentz factor, and the minimum electron energy, have to be large, in order to avoid fast cooling of the electrons \citep{Nakar2017, Iyyani2016, Burgess2016, Burgess2018, Oganesyan2019, Ravasio2019}.  In particular, the emission radii that are found ($\ga 3 \times 10^{16}$ cm) are close to typical values for the deceleration radius\footnote{Large emission radii set limits on the minimum timescale of variations in the light curve, the variabililty time-scale  $t_{\rm var} = R/2c\Gamma^2\, (1+z) \, \sim 5$ s $(\Gamma/300)^{-2} (1+z)$, where $z$ is the redshift. In many of these cases, though,  the time variability constraint is not a problem \citep{Golkhou&Butler2014}. However, in for instance, GRB090926A the high variability of the MeV light curve ($\sim 0.1$ s) suggests that the MeV component should be in a fast cooling regime, which is in contradictions to the observed spectral shape \citep{Yassine2017}.}. These facts again raises the possibility that a detected "prompt" synchrotron emission could actually from the external shocks. 

 In order to remove the obvious ambiguity and unclarity in GRB emission nomenclature, given by the use of the "prompt" and "afterglow" emission phases, 
it might therefore be clearer to instead use the dichotomy between the photosphere (with various degrees of dissipation;  \citet{Rees&Meszaros2005}) and the blastwave (Synchrotron self-Compton emission; \citet{ReesMeszaros1992} or inverse Compton cooling; \citet{Beloborodov2014}).

\subsection{Spectral Evolution During Pulses}

One particularity in our analysis is that we  only studied one spectrum from each burst, namely the narrowest spectrum. This was motivated by the fact that  the spectral shape varies significantly within pulses \citep{Weaton1973, Crider1997, Ryde2019} and therefore the narrowest spectrum is the most constraining for any emission mechanism. Moreover, if the emission mechanism is the same throughout the pulse, then the narrowest spectrum is the  most informative for the pulse emission as a whole. We note that these spectra typically occur close to the pulse peak \citep{Yu2019}, which ensures high signal rates.  

One advantage of such a choice, compared to data samples that include many timebins for each of the studied bursts, is the avoidance of the larger fraction of broad spectra ($\alpha < -2/3$), simply caused by the spectral broadening \citep{Weaton1973}, and that therefore makes model determination ambiguous. A second advantage is that such a choice will not be biased towards bursts with many time bins, which complicates the interpretation of the parameter distributions \citep{Yu2019}.

Within  photospheric emission models the observed broadening of the spectra during a pulse is typically taken as an indication of a varying strength of subphotospheric dissipation \citep[e.g.,][]{Ryde2011, ahlgren2015confronting}. Furthermore, the narrowest spectrum is expected as the photosphere occurs close to the saturation radius, since then dissipation is less likely and at the same time the photospheric emission is the brightest \citep{Ryde2019}.
Therefore, we argue that if a pulse has one spectrum that is convincingly photospheric, then the whole pulse is photospheric.

\section{Conclusions}

We have investigated the narrowness of GRB prompt emission spectra, in order to infer constraints on possible emission mechanisms.
Our sample consists of one spectrum in each of 37 GRB pulses, all selected by having the maximal $\alpha$-value during the pulse, i.e., the narrowest spectrum. The spectra all occur close to the pulse peak and mostly within 10 s of the trigger time. We have shown that  more than half  of these spectra  statistically prefer the non-dissiptive photosphere spectrum (NDP) over the slow-cooled synchrotron spectrum, thereby indicating narrow spectra and  a photospheric origin. Making the natural assumption that the whole pulse due to the same emission mechanism, this conclusion translates to the whole pulse, not only for the narrowest spectrum. 

We also confirm earlier claims in \citet{Acuner2019} that a non-negligible fraction of all spectra ($1/3$), not only statistically prefer, but are also compatible with photospheric emission in a non-dissipative jet. This result is important for two reasons. First, it shows that a noticable fraction of GRB spectra are formed in a region which is not substantially affected by dissipation. This could  be due to the photosphere occurring either close to the saturation radius or in a laminar region of the flow. Second, in the photospheric scenario only a small fraction of spectra are expected to be non-dissipative, since only little disturbances in the flow is needed to change the spectral shape \cite[see, e.g.][]{Peer2006, Ryde2011, Ryde2017}. Therefore, identifying a non-negligible fraction of NDP spectra means that the main part of the remaining spectra should also be photospheric, albeit suffering alterations due to disspation. 

Indeed, among the spectra in our sample there are cases having relatively soft spectra (with $-0.6 \la \alpha \la -0.3$) for which the data still prefer the NDP model.  We have shown that there must be some additional broadening mechanism to explain these spectra, presumably subphotospheric dissipation.
Arguing further, the fact that a broadening mechanism is required for these spectra leads to the natural consequence that a fraction of the rest of the sample, which do not prefer the NDP,  also could be from the photosphere. 
Consequently, there is a natural possibility that a significant majority of the analysed spectra are simply photospheric and only a small fraction are from synchrotron emission:
While a third 
are from non-dissipative photospheres, slightly more than half 
would thus be from dissipative photospheres, leaving only very few 
to be  from synchrotron emission. 
Such a possibility, however,  cannot be firmly established based on the spectral shape alone, as illustrated in Figure \ref{fig:Bjorn}. Other indications, such as the high degree of polarisation that is observed during the prompt phase in some bursts, is a strong indicator for a non-thermal origin of these spectra \citep[e.g., ][]{Fraija2017, Sharma2019}. Moreover, any sample of the prompt emission is most likely contaminated with afterglow emission, which naturally provides synchrotron spectra from the external shock being observed during the "prompt" phase \citep[e.g., ][]{Burgess2016}.




For the model comparison, we have made use of the Bayesian evidences. We have, however, shown that, at least for the spectra in our sample and for the parameter priors we employ, the simper AIC measure gives comparable results. 
Furthermore, we have shown that the traditionally used ``line-of-death'' for synchrotron emission \citep{Preece1998} can be used as a rough discriminator between photospheric and synchrotron spectra. However, the limit is not exact, and $\alpha \ga -0.5$ is safer to use.


Our conclusion is therefore that the initial emission episode in most GRBs is photospheric, be it with or without spectral broadening mechanisms, such as subphotospheric dissipation. However, the actual fraction of bursts with episodes dominated by synchrotron emission during the "prompt" phase is difficult to assess and still needs to be determined, using other observables, such as gamma-ray polarisation. The large fraction of photospheric spectra that we find in this paper, 
is in line with the current paradigm of GRB emission with an efficient photosphere combined with an external blastwave component 
(see review by \citet{Meszaros2019}). 
However, the properties and cause of the subphotospheric dissipation are not firmly established.
This can, though, be investigated by further analysis of observations and numerical simulations of different jet compositions and dynamics.

\acknowledgments

{This research has made use of data obtained through the High Energy Astrophysics Science Archive Research Center Online Service, provided by the NASA/Goddard Space Flight Center. We acknowledge support from the Swedish National Space Agency, from the Swedish Research Council (Vetenskapsr{\aa}det), and from the Swedish Foundation for International Cooperation in Research and Higher Education (STINT). F.R. acknowledge support from  the  G\"oran Gustafsson Foundation for Research in Natural Sciences and Medicine, while A.P. acknowledge support from the EU via the ERC grant O.M.J.}

\appendix

\section{Analysis of the Synthetic and Observed Data}
\label{sec:App_simulations}

By generating data from a selected model or directly from the observed data, we obtain the count spectrum which is the sum of burst flux convolved with the GBM response and the background. These are used to create the Detector Reponse Matrix (DRM). Since this matrix is not square, the equation linking counts to energy spectrum cannot be solved and hence the method of forward folding is required. The model flux vector is folded through the response matrix which gives the model count spectrum. Then, the model count spectrum is compared to the observed counts and a new model flux vector is calculated. Once this iteration leads to satisfactory results, the best fit parameters are obtained.\footnote{See https://fermi.gsfc.nasa.gov/ for a detailed explanation on the analysis of GBM data and the method of forward folding.} Since the GBM data needs the forward folding method to obtain resulting spectra, we compare the generated counts and asses in which ways they deviate from the actual spectra which presents a handy tool for determining good and bad fits from the actual analysis in this paper. As shown in Figures 6 to 9, an ideal fit gives a near perfect one to one relation in the PPC plots and a randomly distributed residual plot whereas a bad fit shows predicted cumulative counts that are outside the 68 \% and 95 \% quantiles for a significant part of the data which looks very different from the one to one relation. Furthermore, the spectrum of such a fit would show systematic differences in the residuals plot.
\subsection{Analysis of the fits from synthetic spectra}
In this section we perform the full statistical analysis for sets of synthetic data, produced from the two theoretical models, NDP and SCS. In such a way we can have clear examples of what good (simulated data fitted by the model it was generated from) and bad fits (simulated data fitted with another model) look like in PPC plots. Additionally, this provides a check on the method used for the analysis, identifying possible coding errors and any faulty selection of prior ranges.

The PPC plots for the analysis on synthetic data are presented in Figs \ref{Fig:Agood} and \ref{Fig:Abad}, which summarises the abovementioned results. 

\begin{figure}
    \centering
    \begin{minipage}{0.45\textwidth}
        \centering
        \includegraphics[width=0.9\textwidth]{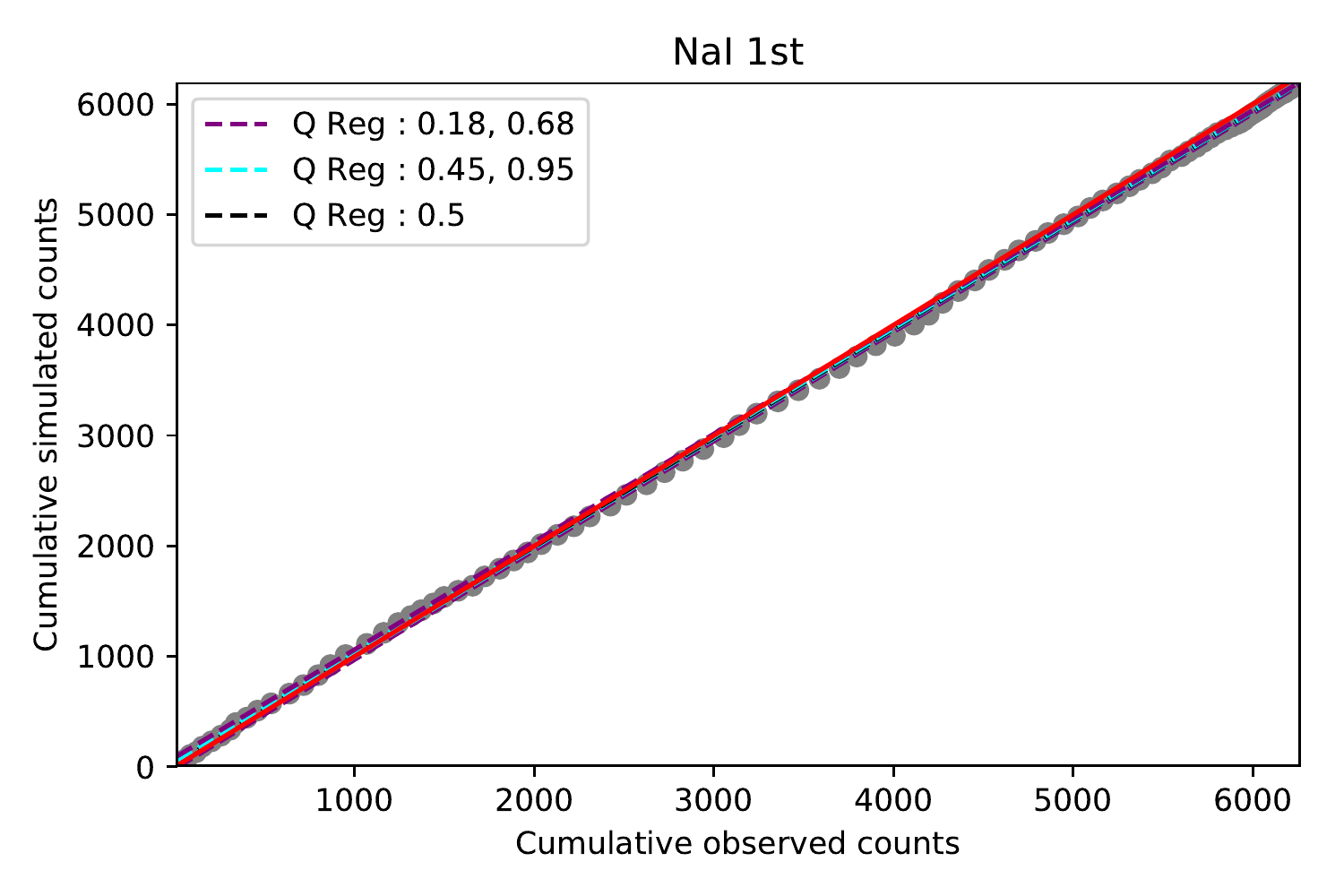} 
    \end{minipage}\hfill
    \begin{minipage}{0.45\textwidth}
        \centering
        \includegraphics[width=0.9\textwidth]{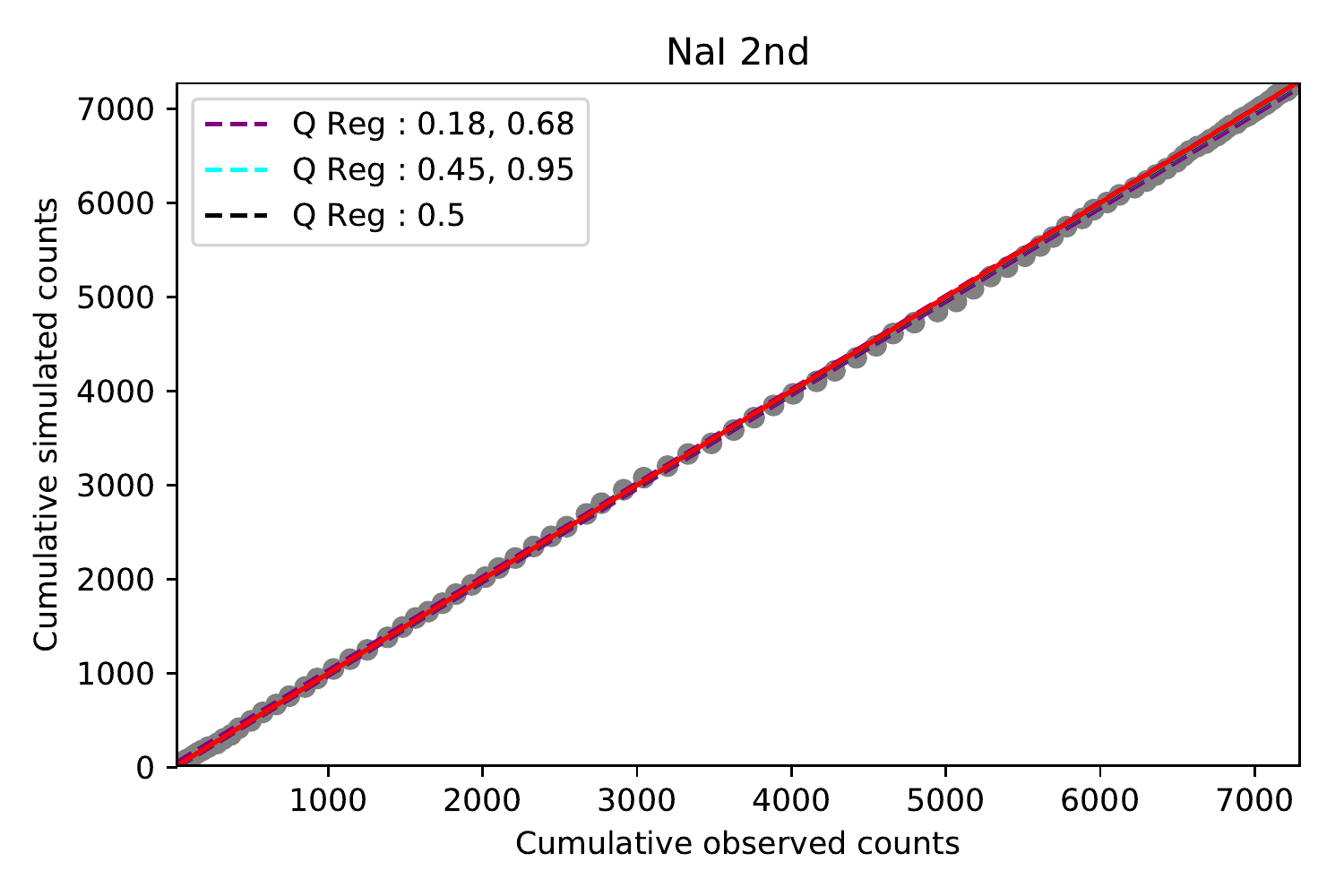} 
    \end{minipage}
\end{figure}

\begin{figure}
    \centering

    \begin{minipage}{0.45\textwidth}
      \centering
        \includegraphics[width=0.9\textwidth]{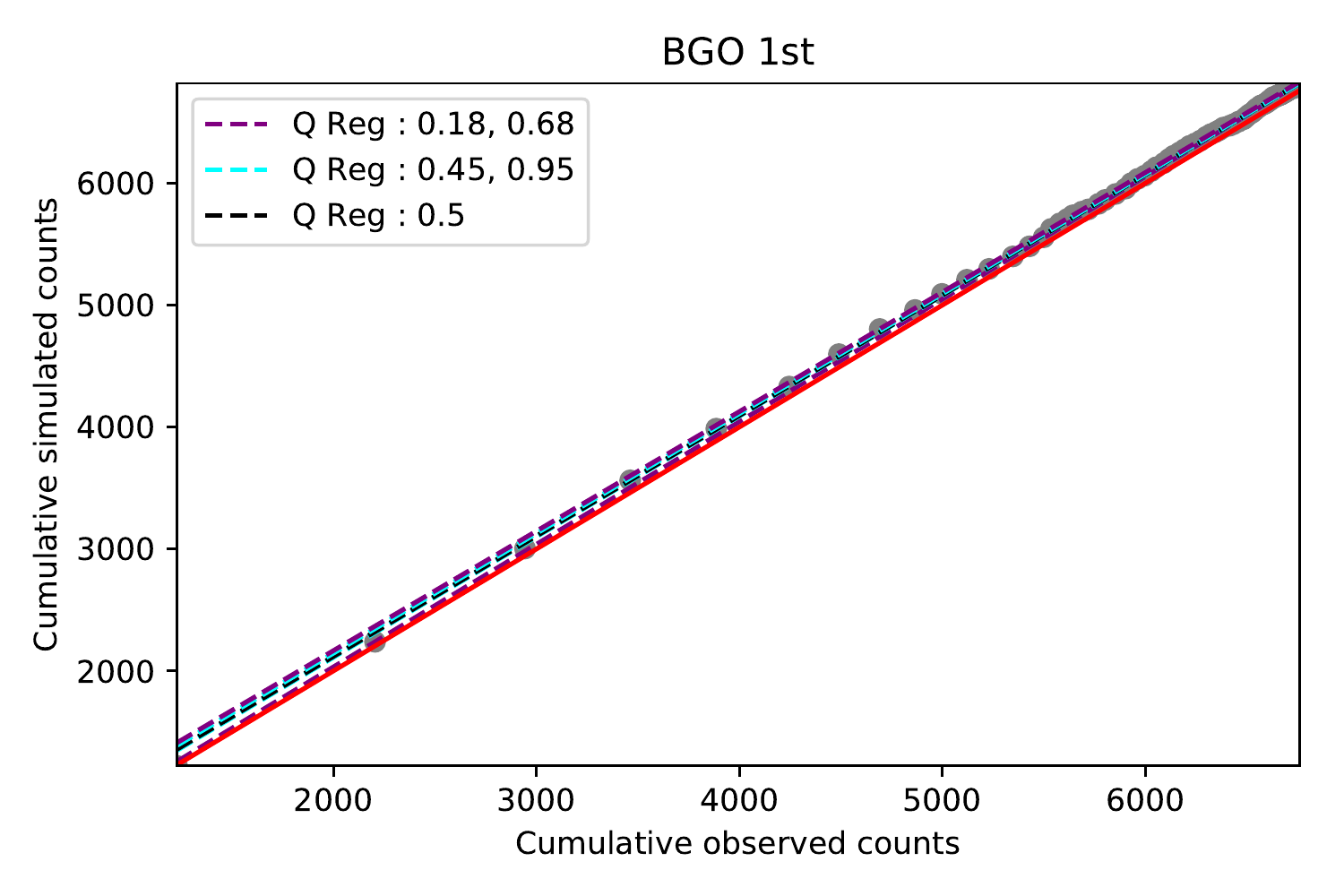} 
    \end{minipage}
\caption{Posterior Predictive Check (PPC) count plots from a well fitted spectrum, with three GBM detectors shown in separate panels. The red line demonstrates the one-to-one relation. In the lower right-hand panel the count spectrum and the best fit model are shown, together with the residuals. The data are from a synthetic NPD spectrum with a SNR=100 and it was fitted with the NPD spectral model. }\label{Fig:Agood}
\end{figure}

\begin{figure}
\gridline{\fig{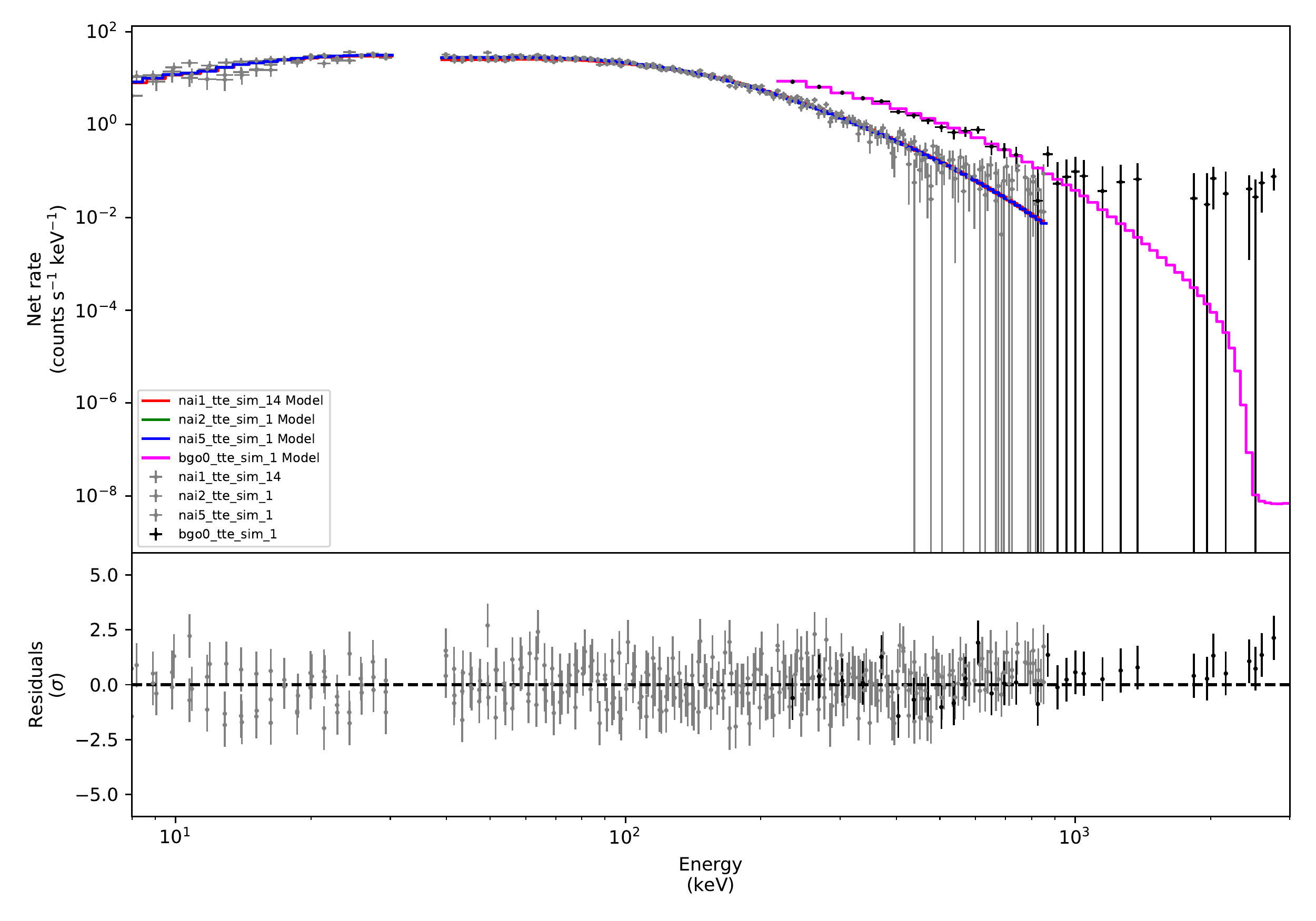}{0.6\textwidth}{}}
\label{Fig:AgoodSpectrum}
\caption{Count spectrum for the NDP fit to the simulated NDP spectrum as explained in Appendix B.}
\end{figure}

\begin{figure}
    \centering
    \begin{minipage}{0.45\textwidth}
        \centering
        \includegraphics[width=0.9\textwidth]{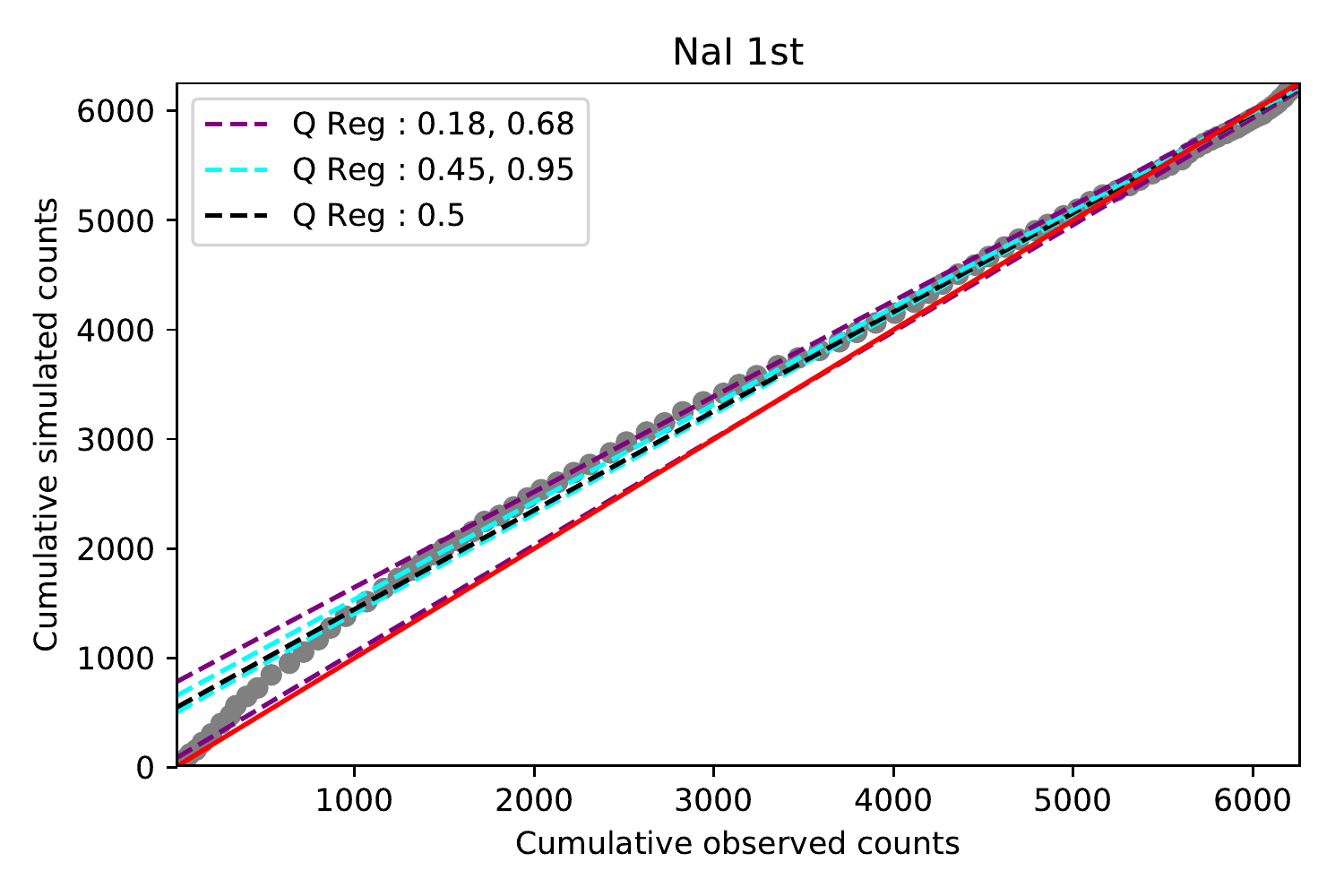} 
    \end{minipage}\hfill
    \begin{minipage}{0.45\textwidth}
        \centering
        \includegraphics[width=0.9\textwidth]{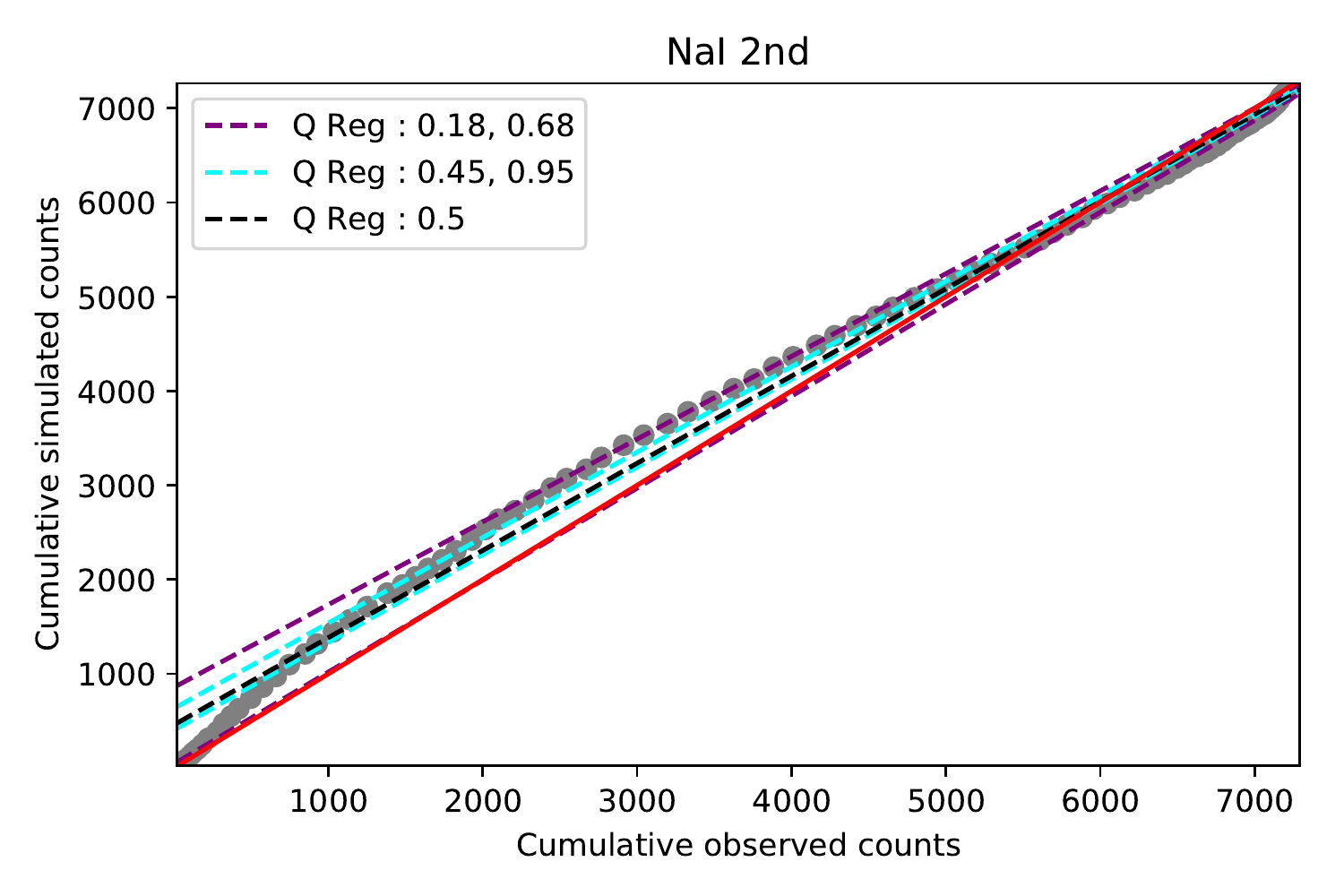} 
    \end{minipage}
\end{figure}

\begin{figure}
    \centering

    \begin{minipage}{0.45\textwidth}
      \centering
        \includegraphics[width=0.9\textwidth]{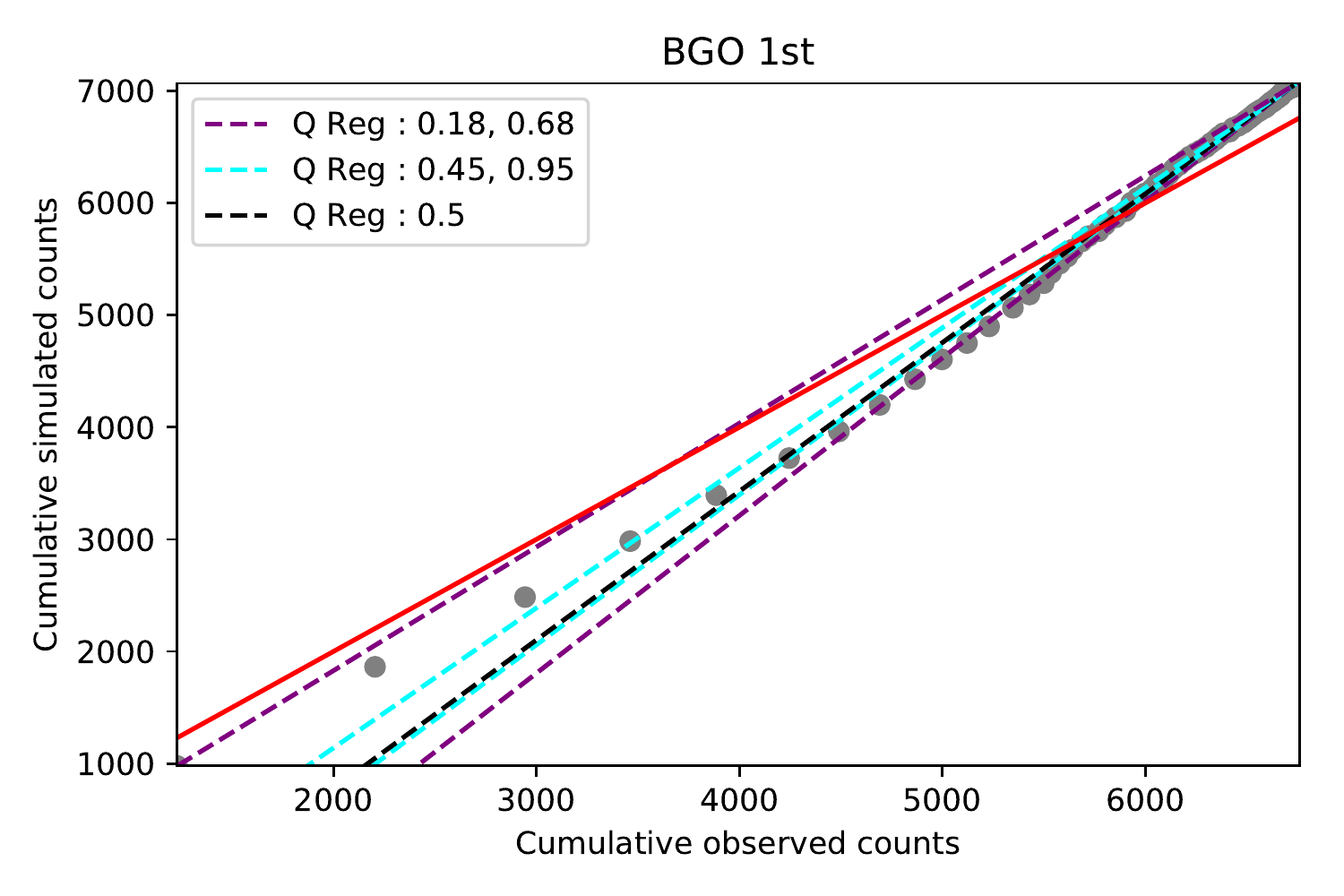} 
    \end{minipage}
\caption{Same as Fig. \ref{Fig:Agood}, but for a badly fitted spectrum. The synthetic data (NDP spectrum, SNR=100) are now fitted with the SCS spectral model, instead. Deviations from the one-to-one lines are apparent.}\label{Fig:Abad}

\end{figure}

\begin{figure}
\gridline{\fig{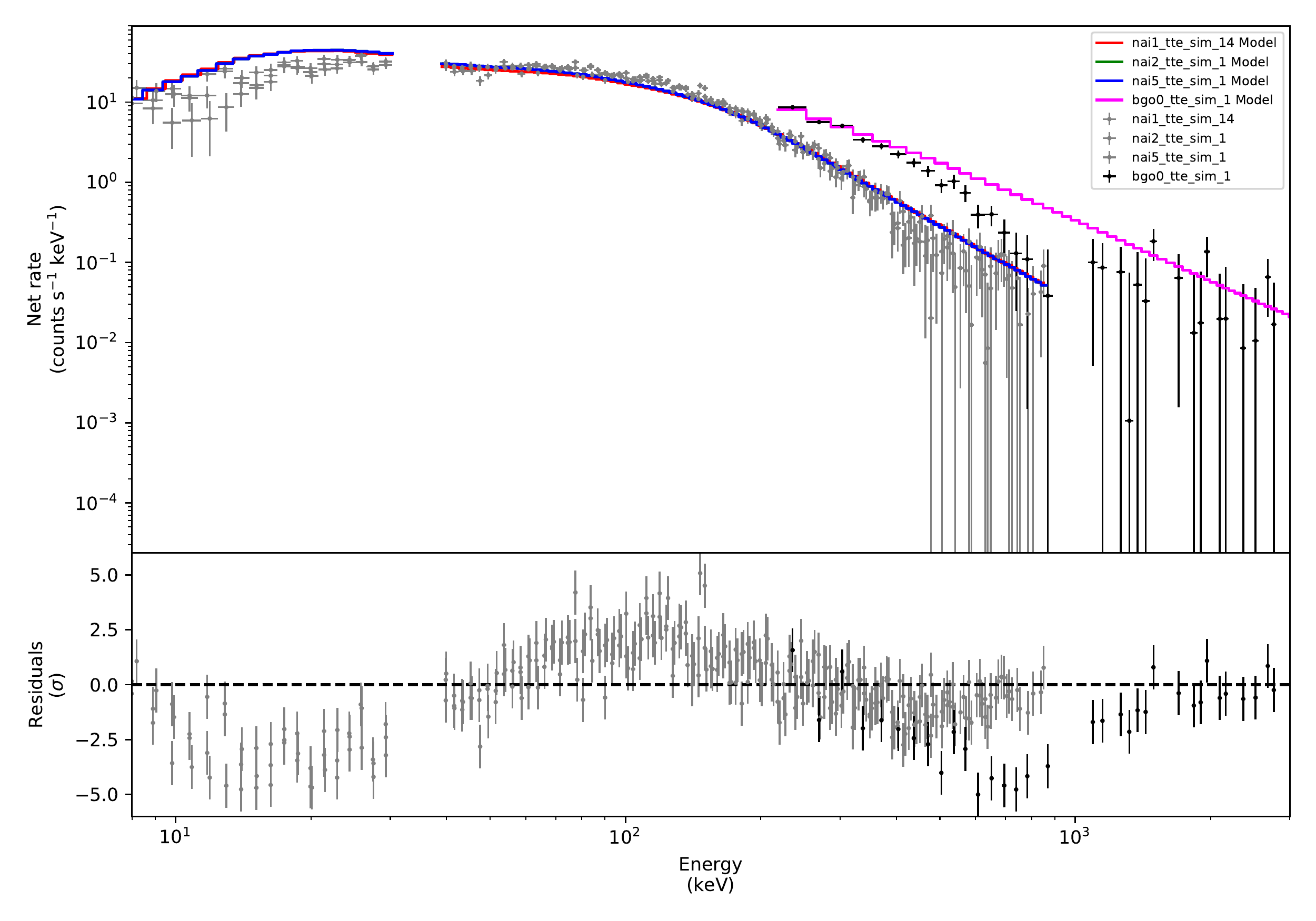}{0.6\textwidth}{}}
\caption{Count spectrum for the SCS fit to the simulated NDP spectrum as explained in Appendix B.}
\label{Fig:AbadSpectrum}
\end{figure}

\subsection{Further Results of the Analysis on the Sample}
\label{sec:App_results}

In Table \ref{tab:appendix} we present more details of the analysis of the 37 spectra in our sample. For every spectrum we present the statistical significance, $S$, of the timebin \citep[as defined in ][]{Vianello2018}, and its time-interval (cf. to \citet{Yu2019}). 
Furthermore, we present the statistical measures AIC \citep{akaike}, BIC \citep{bic}, and DIC \citep{Gelman2014} for the nondissipative photosphere  and the slow cooled synchrotron spectral fits. 

All the fits were initially performed with a maximum likelihood estimation. We used these point estimates as initial starting values for our fits within the Bayesian framework. We find that both sets of fits give consistent estimations within the error bars for all bursts.

\begin{table}[ht]
\centering \label{tab:appendix}
\begin{tabular}{rcr|rc|rrr|rrr}
  \hline
 GRB ID & $S$ & Interval [s] & $\ln Z_{\rm CPL}$&  $\ln Z_{\rm NDP}$-$\ln Z_{\rm CPL}$&$AIC_{\rm NDP}$ & $BIC_{\rm NDP}$  & $DIC_{\rm NDP}$  & $AIC_{\rm SCS}$  & $BIC_{\rm SCS}$  & $DIC_{\rm SCS}$  \\ 

  \hline
  081009140 & 74.27 &   5.65 - 6.35 & -627.58 & -0.30 & 1261.46 & 1268.69 & 1244.84 & 1468.63 & 1475.85 & 1451.94 \\ 
  081125496 & 27.90 &   2.87 - 3.71 & -1404.48 & -1.67 & 2819.19 & 2827.43 & 2803.61 & 2827.77 & 2836.01 & 2811.16 \\ 
081224887 & 48.00 &   0.35 - 1.76  & -1525.13 & -1.73 & 3062.79 & 3071.05 & 3043.57 & 3324.00 & 3332.26 & 3303.06 \\ 
090530760 & 22.28 &   2.89 - 4.98 & -2036.12 & 2.28 & 4071.35 & 4079.61 & 4057.03 & 4136.84 & 4145.09 & 4121.91 \\ 
090620400 & 34.73 &   0.96 - 3.21 & -1571.07 & -0.56 & 3145.59 & 3153.84 & 3131.48 & 3339.58 & 3347.83 & 3324.65 \\ 
090626189 & 39.26 &   34.43 - 34.64 & -246.09 & -14.53 & 532.97 & 541.25 & 513.37 & 523.63 & 531.90 & 503.24 \\ 
090719063 & 38.25 &   0.72 - 1.38 & -723.63 & -2.69 & 1460.56 & 1468.83 & 1442.52 & 1687.95 & 1696.22 & 1668.87 \\ 
090804940 & 22.01 &   0.62 - 0.97  & -419.61 & 0.16 & 844.26 & 852.51 & 829.88 & 886.80 & 895.06 & 871.73 \\ 
090820027 & 40.42 &   31.16 - 31.39  & -117.61 & -5.07 & 255.26 & 263.53 & 235.36 & 293.83 & 302.11 & 273.45 \\ 
100122616 & 39.79 &   21.97 - 22.59 & -1243.02 & -50.79 & 2591.73 & 2599.98 & 2576.58 & 2479.29 & 2487.54 & 2463.64 \\ 
100528075 & 31.61 &   11.78 - 12.73 & -1021.46 & -44.79 & 2136.75 & 2144.99 & 2122.51 & 2048.11 & 2056.35 & 2033.16 \\ 
100612726 & 33.53 &   2.86 - 3.30 & -201.02 & 1.24 & 405.56 & 413.81 & 389.74 & 478.63 & 486.89 & 462.15 \\ 
100707032 & 34.95 &   0.36 - 0.70  & -1860.56 & -31.61 & 3793.70 & 3801.96 & 3771.92 & 5824.53 & 5832.78 & 5801.01 \\ 
101126198 & 47.68 &   9.75 - 11.75 & -2228.31 & -60.99 & 4581.05 & 4589.29 & 4567.79 & 4453.44 & 4461.68 & 4439.61 \\ 
110817191 & 20.48 &   0.43 - 0.81  & -864.98 & -4.76 & 1748.78 & 1757.02 & 1731.16 & 1752.76 & 1761.01 & 1734.46 \\ 
110721200 & 75.54 &   1.96 - 2.97  & -1186.93 & -262.92 & 2905.21 & 2913.46 & 2887.34 & 2325.19 & 2333.44 & 2306.59 \\ 
110920546 & 19.85 &   89.25 - 105.72  & -3510.94 & -10.52 & 7041.88 & 7050.14 & 7030.70 & 7221.74 & 7230.00 & 7210.21 \\ 
111017657 & 36.35 &   3.60 - 4.71 & -1171.10 & -67.19 & 2483.96 & 2492.22 & 2466.65 & 2352.18 & 2360.44 & 2333.55 \\ 
120919309 & 56.61 &   2.68 - 3.53 & -913.47 & -52.79 & 1938.67 & 1946.95 & 1921.16 & 1860.43 & 1868.71 & 1842.12 \\ 
130305486 & 37.76 &   4.45 - 5.01 & -1034.81 & -9.88 & 2098.87 & 2107.13 & 2079.91 & 2155.58 & 2163.84 & 2135.68 \\ 
130815660 & 42.16 &   33.01 - 33.54 & -606.07 & -6.80 & 1230.63 & 1238.89 & 1215.21 & 1232.96 & 1241.21 & 1216.99 \\ 
130612456 & 58.25 &   1.54 - 2.41   &-1063.38 & -1063.01 & -59.33 & 2249.93 & 2258.18 & 2233.39 & 2237.61 & 2245.92  \\ 
130614997 & 36.80 &   2.03 - 3.20  & -1001.43 & -1000.59 & -39.61 & 2083.08 & 2091.33 & 2069.70 & 2002.81 & 2011.06 \\ 
140508128 & 74.19 &   4.71 - 5.03  & -430.39 & -70.53 & 1013.42 & 1020.64 & 992.03 & 895.15 & 902.37 & 873.08 \\ 
141028455 & 30.74 &   10.67 - 10.90 & -1350.68 & -19.50 & 2747.10 & 2755.35 & 2731.02 & 2702.35 & 2710.61 & 2685.19 \\ 
141205763 & 42.91 &   1.44 - 2.32 & -1439.90 & -15.00 & 2914.97 & 2923.24 & 2899.16 & 2914.32 & 2922.58 & 2898.04 \\ 
150213001 & 146.15 &   2.23 - 2.52 & -605.40 & -215.54 & 1650.45 & 1658.69 & 1629.17 & 1407.40 & 1415.64 & 1386.21 \\ 
150306993 & 23.33 &   0.03 - 1.97  & 1962.71 & 1969.94 & 1946.60 & 2030.78 & 2038.02 & 2013.95&2220.55&1988.99 \\ 
150314205 & 47.82 &   1.09 - 1.27 & -241.51 & 3.82 & 487.31 & 495.56 & 466.27 & 653.69 & 662.00 & 631.48 \\ 
150510139 & 39.91 &   0.90 - 1.11  & -30.21 & -40.53 & 155.66 & 163.93 & 133.50 & 151.63 & 159.96 & 128.37 \\ 
150902733 & 92.69 &   8.93 - 9.41  & -807.25 & -6.39 & 1637.90 & 1646.16 & 1616.19 & 2065.75 & 2074.01 & 2043.37 \\ 
151021791 & 29.06 &   0.15 - 0.62  & -500.53 & 2.96 & 1003.77 & 1012.03 & 986.12 & 1086.83 & 1095.09 & 1067.94 \\ 
160215773 & 20.00 &   174.29 - 175.91 & -2301.15 & -96.55 & 4798.88 & 4807.15 & 0.00 & 4529.39 & 4537.65 & 0.00 \\ 
160530667 & 174.09 &  4.56 - 5-51 & -1527.18 & -196.69 & 3455.36 & 3463.64 & 3434.23 & 3594.53 & 3602.81 & 3572.76 \\ 
160910722 & 22.30 &  7.00 - 7.33 & -291.83 & -1.25 & 598.38 & 606.64 & 578.26 & 680.79 & 689.05 & 659.77 \\ 
161004964 & 22.61 &  1.02 - 2.69  & -1575.92 & 2.63 & 3151.06 & 3159.31 & 3136.86 & 3193.09 & 3201.35 & 3177.82 \\  
170114917 & 50.30 &   1.99 - 2.54  & -877.30 & -9.24 & 1780.93 & 1789.20 & 1762.93 & 1812.19 & 1820.46 & 1793.09 \\
   \hline
\end{tabular}
\caption{Statistical significance $S$ \citep{Vianello2018} and the time intervals used for the analysed sample are presented. Additionally, the evidences for CPL fits, the difference between CPL and NDP evidences as well as the corresponding information criteria for NDP and SCS fits are given. }
\end{table}

\subsection{Detailed Analysis Results of Two Example Spectra}
\label{sec:App_Examples}

As part of the analysis the fits were judged based on a number of characteristics.
First,  the spectral fit for the various models is displayed on the observed counts and shown together with the residuals. A wavy structure of the residuals indicate a bad fit.

Second, we present the PPC plots, which again gives an indication of how well the model fits the data.

Third, the Bayesian analysis produces corner plots of the posterior distributions of the parameters.
These fits are considered good if a Gaussian like distribution is obtained for the posteriors.

To illustrate this proceedure, we present the detailed analysis on GRB150314 (Fig.\ref{Fig:A110920}) and GRB150902  (Fig. \ref{Fig:A150902}) as examples. We again show the same plots as in Figure \ref{Fig:Agood}, but now these figures are on the real data from GBM. For GRB150314 we see a perfect one to one relation for the PPC plots and a random distribution of the residuals in the spectral fit which both indicate a good fit. For GRB150902 we see PPC plots that do not match with the one to one PPC line and the spectrum residuals are showing systematic differences which indicate a bad fit. 

For both of these fits we have Gaussian like posterior distributions which shows the fits have converged properly. However, the ability of the two different models to describe the data is very different as seen from the PPC plots and the residuals in the spectra.


\begin{figure}
    \centering
    \begin{minipage}{0.45\textwidth}
        \centering
        \includegraphics[width=0.9\textwidth]{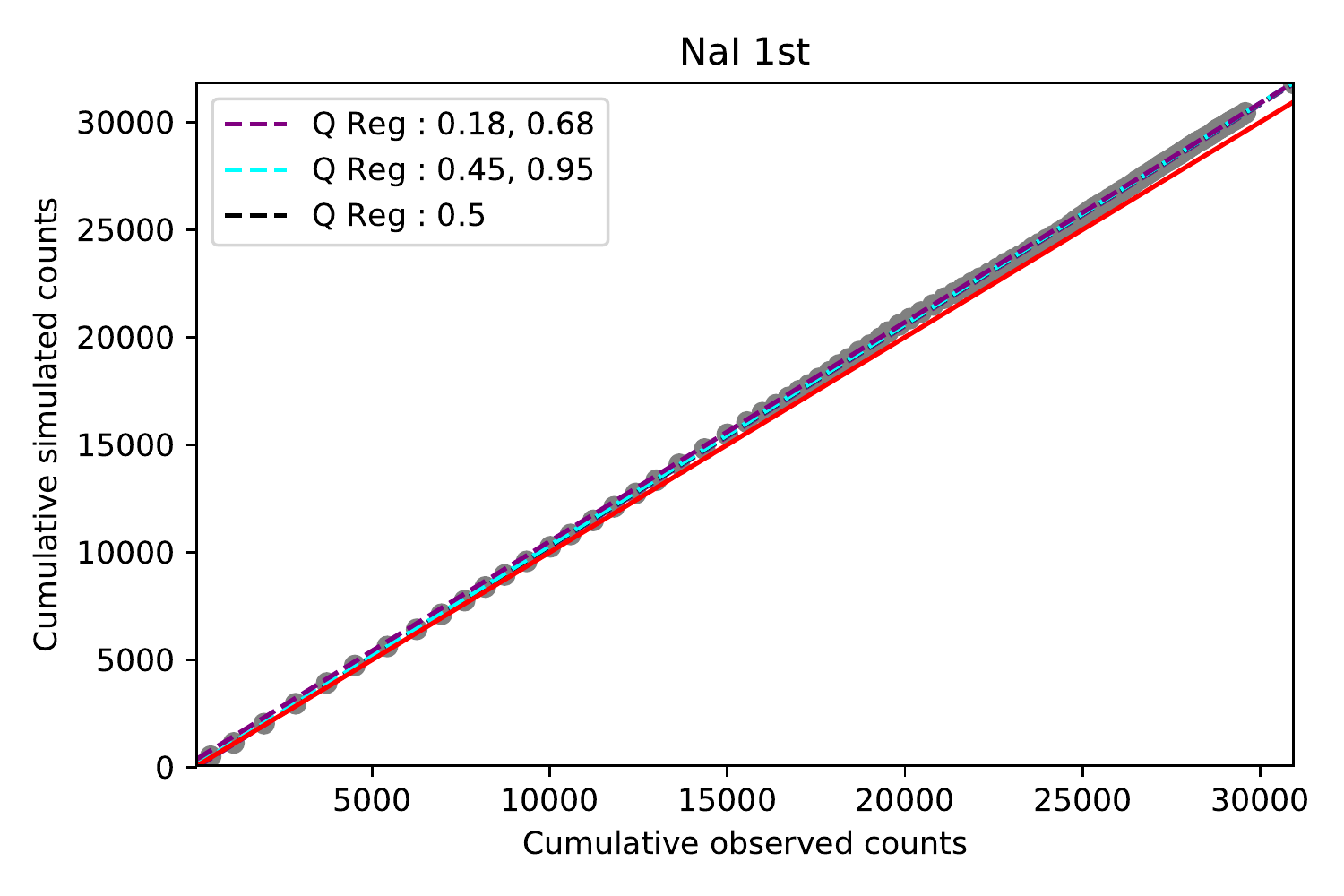} 
    \end{minipage}\hfill
    \begin{minipage}{0.45\textwidth}
        \centering
        \includegraphics[width=0.9\textwidth]{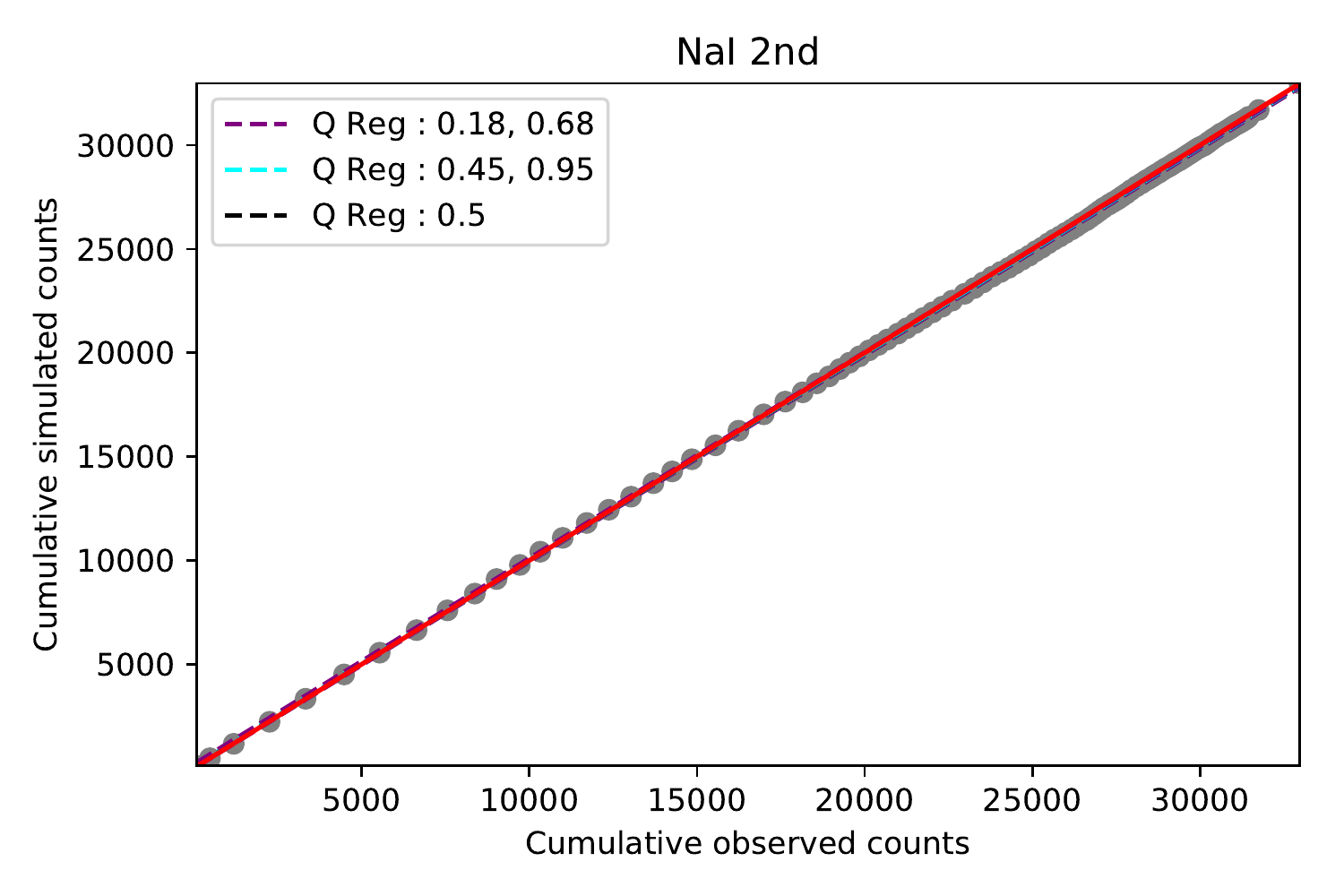} 
    \end{minipage}
\end{figure}

\begin{figure}
    \centering

    \begin{minipage}{0.45\textwidth}
      \centering
        \includegraphics[width=0.9\textwidth]{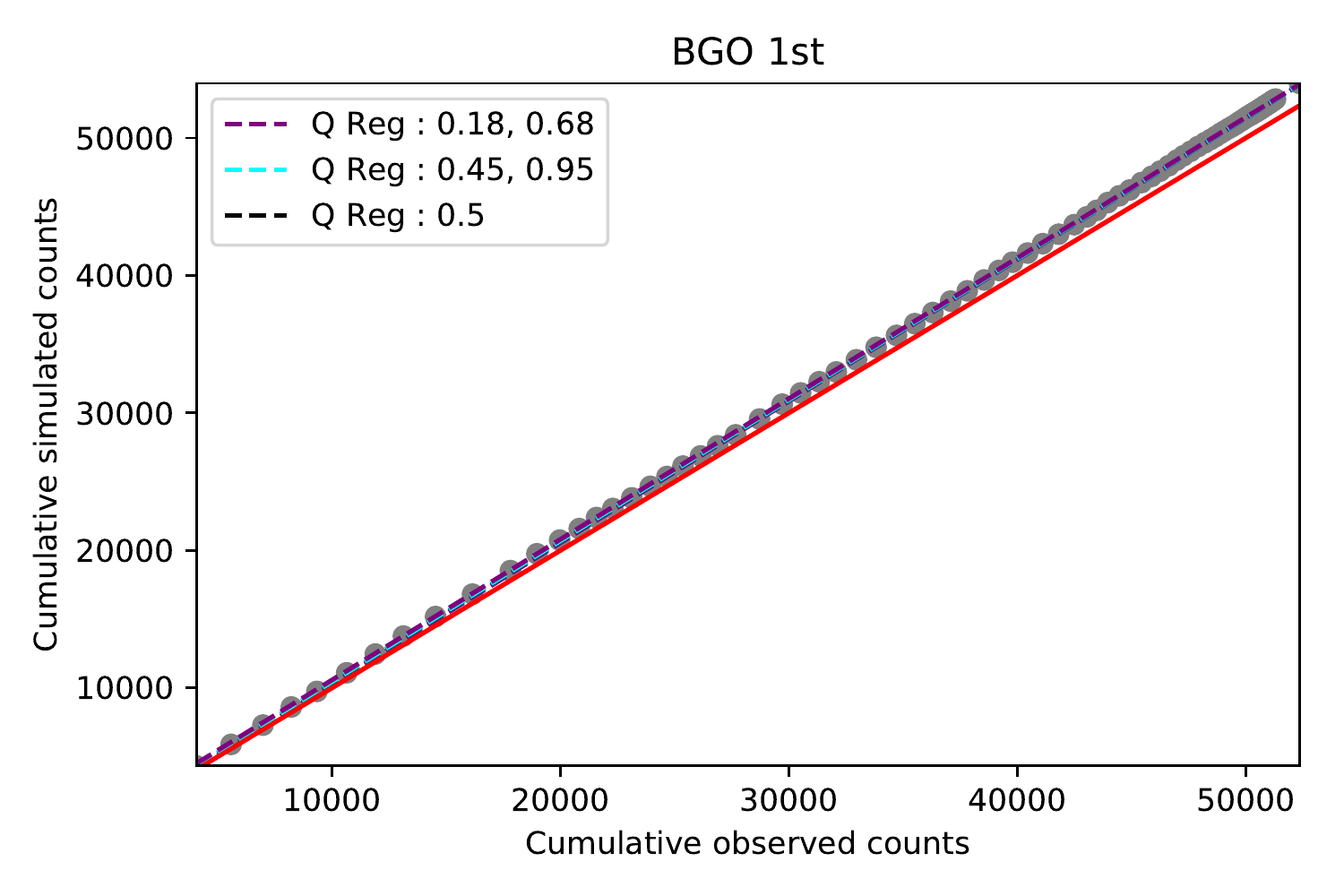} 
    \end{minipage}
\caption{Same as Fig. \ref{Fig:Agood}, but for a well fitted observed GBM spectrum. GRB150314205, with SNR=20, is given as an example. The data are fitted with a NDP model.}\label{Fig:A110920}

\end{figure}

\begin{figure}
\gridline{\fig{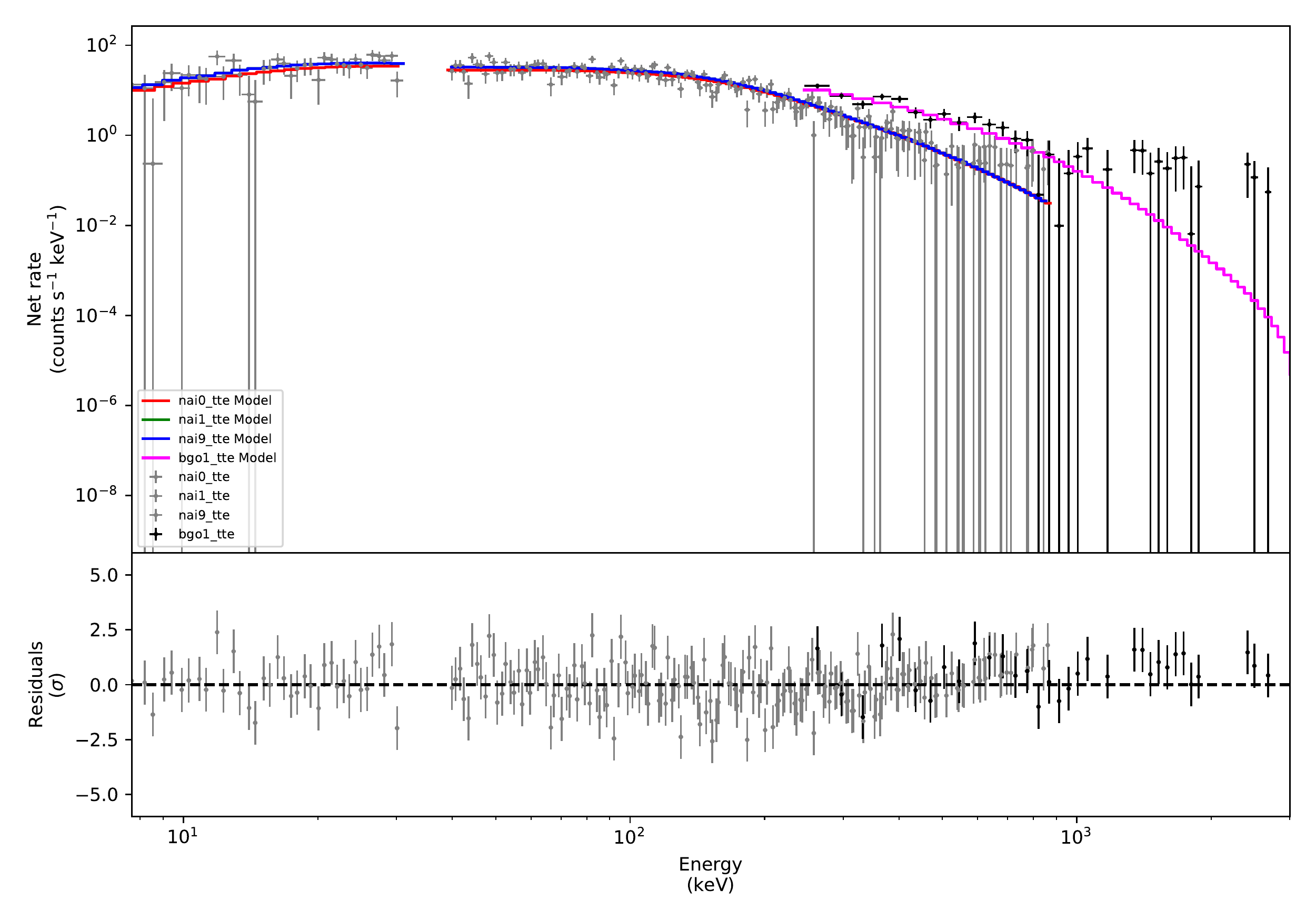}{0.6\textwidth}{}}
\caption{Count spectrum for the NDP fit to GRB150314 as explained in Appendix C.1.}
\label{Fig:A110920Spectrum}
\end{figure}

\begin{figure}
    \centering
    \begin{minipage}{0.45\textwidth}
        \centering
        \includegraphics[width=0.9\textwidth]{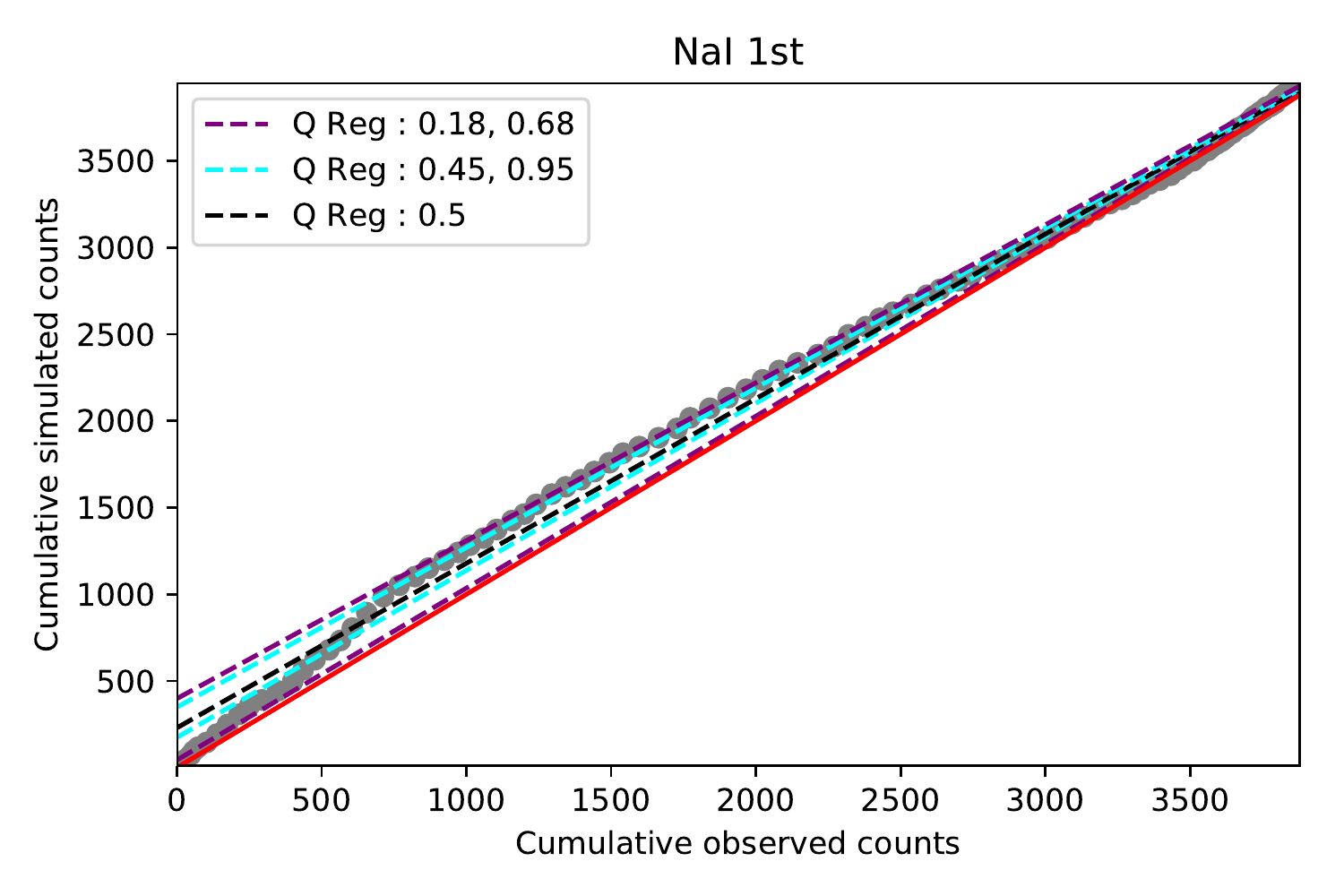} 
    \end{minipage}\hfill
    \begin{minipage}{0.45\textwidth}
        \centering
        \includegraphics[width=0.9\textwidth]{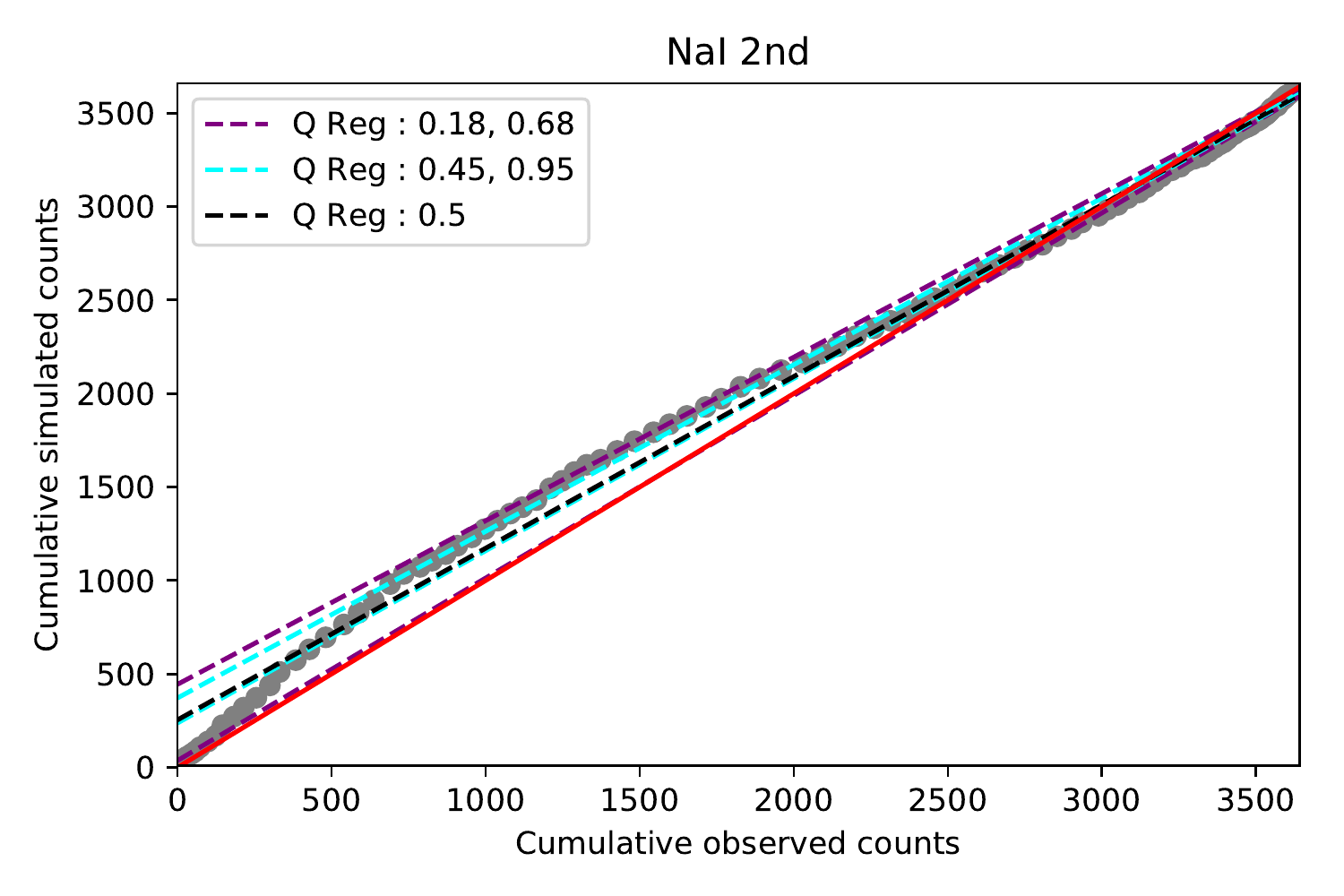} 
    \end{minipage}
\end{figure}

\begin{figure}
    \centering

    \begin{minipage}{0.45\textwidth}
      \centering
        \includegraphics[width=0.9\textwidth]{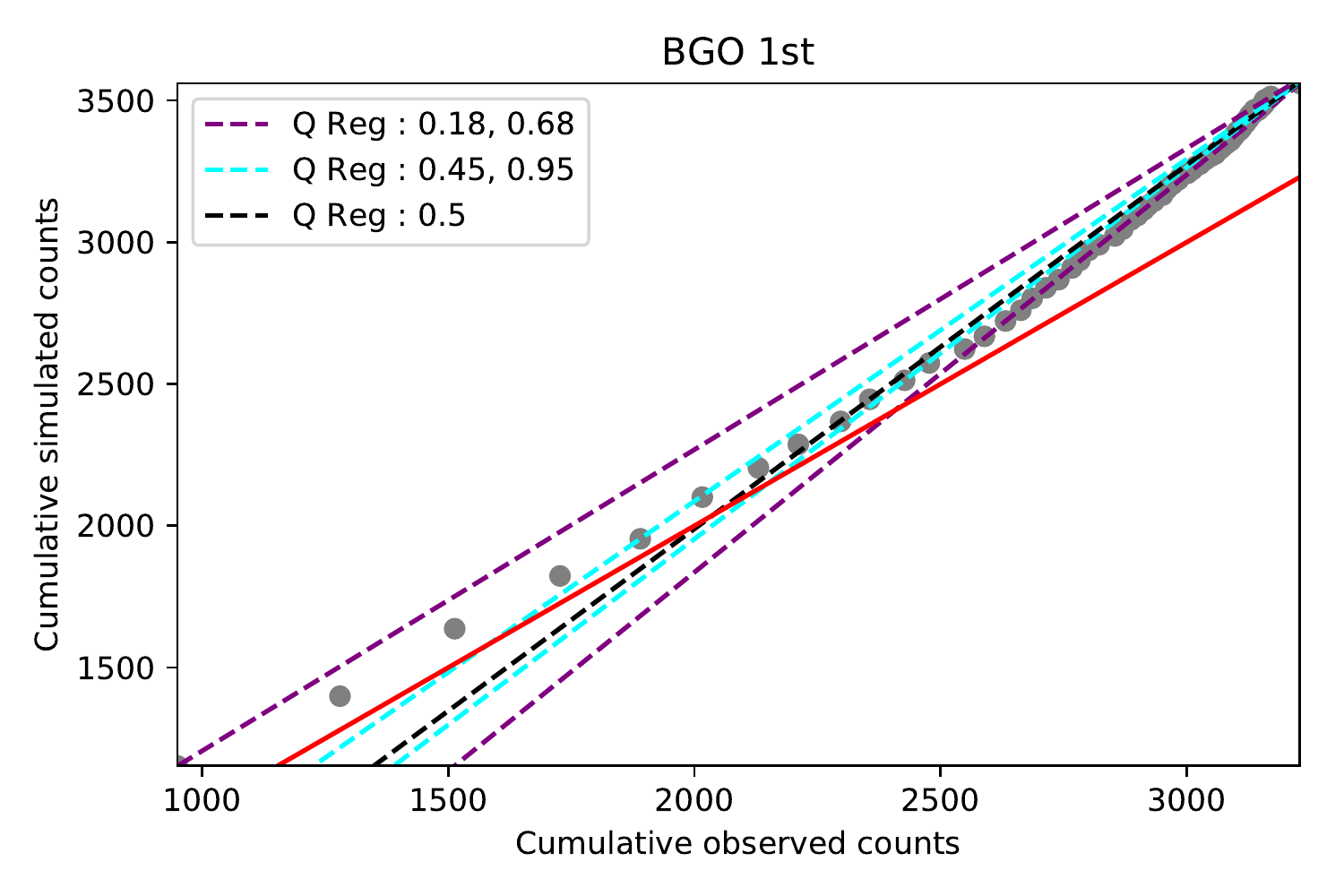} 
    \end{minipage}

\caption{Same as Fig. \ref{Fig:Agood}, but for a badly fitted observed GBM spectrum. GRB150902A, with SNR=93, is given as an example. The data are fitted with a SCS model.}\label{Fig:A150902}

\end{figure}
\begin{figure}
\gridline{\fig{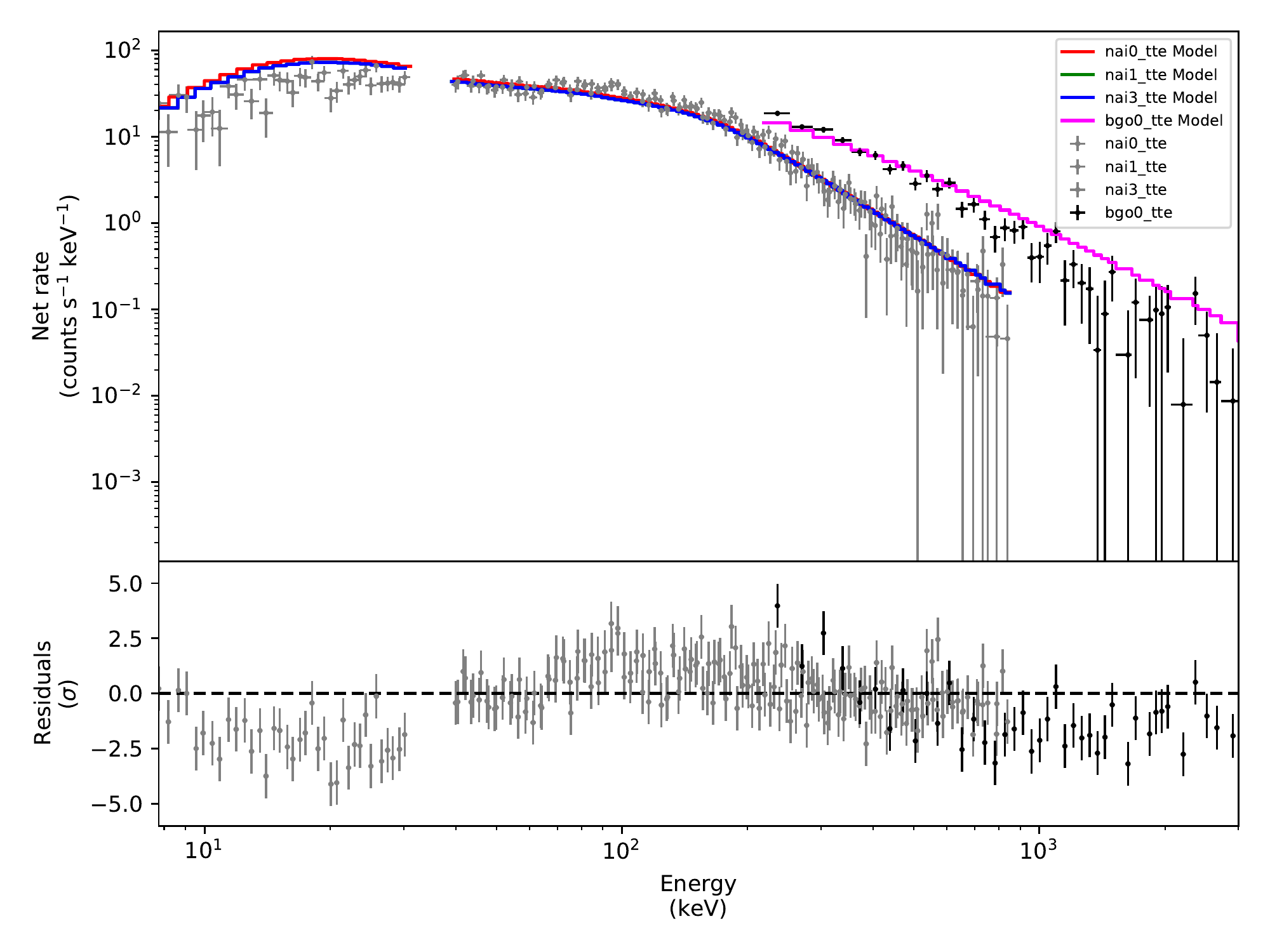}{0.6\textwidth}{}}
\caption{Count spectrum for the SCS fit to GRB150902A as explained in Appendix B.}

\label{fig:A150902Spectrum}
\end{figure}

\bibliographystyle{aasjournal}
\bibliography{ref2017}

\begin{thebibliography}{}
\expandafter\ifx\csname natexlab\endcsname\relax\def\natexlab#1{#1}\fi
\providecommand{\url}[1]{\href{#1}{#1}}

\bibitem[{{Ackermann} {et~al.}(2010){Ackermann}, {Asano}, {Atwood}, {Axelsson},
  {Baldini}, {Ballet}, {Barbiellini}, {Baring}, {Bastieri}, {Bechtol},
  {Bellazzini}, {Berenji}, {Bhat}, {Bissaldi}, {Blandford}, {Bloom},
  {Bonamente}, {Borgland}, {Bouvier}, {Bregeon}, {Brez}, {Briggs}, {Brigida},
  {Bruel}, {Buson}, {Caliandro}, {Cameron}, {Caraveo}, {Carrigan}, {Casand
  jian}, {Cecchi}, {{\c{C}}elik}, {Charles}, {Chiang}, {Ciprini}, {Claus},
  {Cohen-Tanugi}, {Connaughton}, {Conrad}, {Dermer}, {de Palma}, {Dingus},
  {Silva}, {Drell}, {Dubois}, {Dumora}, {Farnier}, {Favuzzi}, {Fegan}, {Finke},
  {Focke}, {Frailis}, {Fukazawa}, {Fusco}, {Gargano}, {Gasparrini}, {Gehrels},
  {Germani}, {Giglietto}, {Giordano}, {Glanzman}, {Godfrey}, {Granot},
  {Grenier}, {Grondin}, {Grove}, {Guiriec}, {Hadasch}, {Harding}, {Hays},
  {Horan}, {Hughes}, {J{\'o}hannesson}, {Johnson}, {Kamae}, {Katagiri},
  {Kataoka}, {Kawai}, {Kippen}, {Kn{\"o}dlseder}, {Kocevski}, {Kouveliotou},
  {Kuss}, {Lande}, {Latronico}, {Lemoine-Goumard}, {Llena Garde}, {Longo},
  {Loparco}, {Lott}, {Lovellette}, {Lubrano}, {Makeev}, {Mazziotta}, {McEnery},
  {McGlynn}, {Meegan}, {M{\'e}sz{\'a}ros}, {Michelson}, {Mitthumsiri},
  {Mizuno}, {Moiseev}, {Monte}, {Monzani}, {Moretti}, {Morselli}, {Moskalenko},
  {Murgia}, {Nakajima}, {Nakamori}, {Nolan}, {Norris}, {Nuss}, {Ohno},
  {Ohsugi}, {Omodei}, {Orlando}, {Ormes}, {Ozaki}, {Paciesas}, {Paneque},
  {Panetta}, {Parent}, {Pelassa}, {Pepe}, {Pesce-Rollins}, {Piron}, {Preece},
  {Rain{\`o}}, {Rando}, {Razzano}, {Razzaque}, {Reimer}, {Ritz}, {Rodriguez},
  {Roth}, {Ryde}, {Sadrozinski}, {Sander}, {Scargle}, {Schalk}, {Sgr{\`o}},
  {Siskind}, {Smith}, {Spandre}, {Spinelli}, {Stamatikos}, {Stecker},
  {Strickman}, {Suson}, {Tajima}, {Takahashi}, {Takahashi}, {Tanaka}, {Thayer},
  {Thayer}, {Thompson}, {Tibaldo}, {Toma}, {Torres}, {Tosti}, {Tramacere},
  {Uchiyama}, {Uehara}, {Usher}, {van der Horst}, {Vasileiou}, {Vilchez},
  {Vitale}, {von Kienlin}, {Waite}, {Wang}, {Wilson-Hodge}, {Winer}, {Wu},
  {Yamazaki}, {Yang}, {Ylinen}, \& {Ziegler}}]{090510}
{Ackermann}, M., {Asano}, K., {Atwood}, W.~B., {et~al.} 2010, \apj, 716, 1178

\bibitem[{{Acuner} {et~al.}(2019){Acuner}, {Ryde}, \& {Yu}}]{Acuner2019}
{Acuner}, Z., {Ryde}, F., \& {Yu}, H.-F. 2019, Monthly Notices of the Royal
  Astronomical Society, 487, 5508

\bibitem[{{Aharonian} {et~al.}(2010){Aharonian}, {Kelner}, \&
  {Prosekin}}]{NaimaSynch}
{Aharonian}, F.~A., {Kelner}, S.~R., \& {Prosekin}, A.~Y. 2010, Physical Review
  D, 82, 043002

\bibitem[{{Ahlgren} {et~al.}(2019){Ahlgren}, {Larsson}, {Ahlberg}, {Lundman},
  {Ryde}, \& {Pe'er}}]{Ahlgren2019}
{Ahlgren}, B., {Larsson}, J., {Ahlberg}, E., {et~al.} 2019, \mnras, 485, 474

\bibitem[{Ahlgren {et~al.}(2015)Ahlgren, Larsson, Nymark, Ryde, \&
  Pe'er}]{ahlgren2015confronting}
Ahlgren, B., Larsson, J., Nymark, T., Ryde, F., \& Pe'er, A. 2015, Monthly
  Notices of the Royal Astronomical Society: Letters, 454, L31

\bibitem[{{Ajello} {et~al.}(2019{\natexlab{a}}){Ajello}, {Arimoto}, {Axelsson},
  {Baldini}, {Barbiellini}, {Bastieri}, {Bellazzini}, {Berretta}, {Bissaldi},
  {Blandford}, {Bonino}, {Bottacini}, {Bregeon}, {Bruel}, {Buehler}, {Burns},
  {Buson}, {Cameron}, {Caputo}, {Caraveo}, {Cavazzuti}, {Chen}, {Chiaro},
  {Ciprini}, {Cohen-Tanugi}, {Costantin}, {Cutini}, {D'Ammando}, {DeKlotz}, {de
  la Torre Luque}, {de Palma}, {Desai}, {Di Lalla}, {Di Venere}, {Fana
  Dirirsa}, {Fegan}, {Franckowiak}, {Fukazawa}, {Funk}, {Fusco}, {Gargano},
  {Gasparrini}, {Giglietto}, {Gill}, {Giordano}, {Giroletti}, {Granot},
  {Green}, {Grenier}, {Grondin}, {Guiriec}, {Hays}, {Horan}, {J{\'o}hannesson},
  {Kocevski}, {Kovac'evic'}, {Kuss}, {Larsson}, {Latronico}, {Lemoine-Goumard},
  {Li}, {Liodakis}, {Longo}, {Loparco}, {Lovellette}, {Lubrano}, {Maldera},
  {Malyshev}, {Manfreda}, {Mart{\'\i}-Devesa}, {Mazziotta}, {McEnery}, {Mereu},
  {Meyer}, {Michelson}, {Mitthumsiri}, {Mizuno}, {Monzani}, {Moretti},
  {Morselli}, {Moskalenko}, {Negro}, {Nuss}, {Omodei}, {Orienti}, {Orlando},
  {Palatiello}, {Paliya}, {Paneque}, {Pei}, {Persic}, {Pesce-Rollins},
  {Petrosian}, {Piron}, {Poon}, {Porter}, {Principe}, {Racusin}, {Rain{\`o}},
  {Rando}, {Rani}, {Razzano}, {Razzaque}, {Reimer}, {Reimer}, {Ryde}, {Saz
  Parkinson}, {Serini}, {Sgr{\`o}}, {Siskind}, {Spandre}, {Spinelli}, {Tajima},
  {Takagi}, {Takahashi}, {Tak}, {Thayer}, {Thompson}, {Torres}, {Troja},
  {Valverde}, {Van Klaveren}, {Wood}, {Yassine}, {Zaharijas}, {Bhat}, {Briggs},
  {Cleveland }, {Giles}, {Goldstein}, {Hui}, {Malacaria}, {Preece}, {Roberts},
  {Veres}, {von Kienlin}, {Cenko}, {O'Brien}, {Beardmore}, {Lien}, {Osborne},
  {Tohuvavohu}, {D'Elia}, {D'A}, {Perri}, {Gropp}, {Klingler}, {Capalbi},
  {Tagliaferri}, \& {Stamatikos}}]{Ajello190114C}
{Ajello}, M., {Arimoto}, M., {Axelsson}, M., {et~al.} 2019{\natexlab{a}}, arXiv
  e-prints, arXiv:1909.10605

\bibitem[{{Ajello} {et~al.}(2019{\natexlab{b}}){Ajello}, {Arimoto}, {Axelsson},
  {Baldini}, {Barbiellini}, {Bastieri}, {Bellazzini}, {Bhat}, {Bissaldi},
  {Blandford}, {Bonino}, {Bonnell}, {Bottacini}, {Bregeon}, {Bruel}, {Buehler},
  {Cameron}, {Caputo}, {Caraveo}, {Cavazzuti}, {Chen}, {Cheung}, {Chiaro},
  {Ciprini}, {Costantin}, {Crnogorcevic}, {Cutini}, {Dainotti}, {D'Ammand o},
  {de la Torre Luque}, {de Palma}, {Desai}, {Desiante}, {Di Lalla}, {Di
  Venere}, {Fana Dirirsa}, {Fegan}, {Franckowiak}, {Fukazawa}, {Funk}, {Fusco},
  {Gargano}, {Gasparrini}, {Giglietto}, {Giordano}, {Giroletti}, {Green},
  {Grenier}, {Grove}, {Guiriec}, {Hays}, {Hewitt}, {Horan}, {J{\'o}hannesson},
  {Kocevski}, {Kuss}, {Latronico}, {Li}, {Longo}, {Loparco}, {Lovellette},
  {Lubrano}, {Maldera}, {Manfreda}, {Mart{\'\i}-Devesa}, {Mazziotta}, {Mereu},
  {Meyer}, {Michelson}, {Mirabal}, {Mitthumsiri}, {Mizuno}, {Monzani},
  {Moretti}, {Morselli}, {Moskalenko}, {Negro}, {Nuss}, {Ohno}, {Omodei},
  {Orienti}, {Orlando}, {Palatiello}, {Paliya}, {Paneque}, {Persic},
  {Pesce-Rollins}, {Petrosian}, {Piron}, {Poolakkil}, {Poon}, {Porter},
  {Principe}, {Racusin}, {Rain{\`o}}, {Rando}, {Razzano}, {Razzaque}, {Reimer},
  {Reimer}, {Reposeur}, {Ryde}, {Serini}, {Sgr{\`o}}, {Siskind}, {Sonbas},
  {Spandre}, {Spinelli}, {Suson}, {Tajima}, {Takahashi}, {Tak}, {Thayer},
  {Torres}, {Troja}, {Valverde}, {Veres}, {Vianello}, {von Kienlin}, {Wood},
  {Yassine}, {Zhu}, \& {Zimmer}}]{Ajello2019}
---. 2019{\natexlab{b}}, \apj, 878, 52

\bibitem[{{Akaike}(1974)}]{akaike}
{Akaike}, H. 1974, IEEE Transactions on Automatic Control, 19, 716

\bibitem[{{Arnaud} {et~al.}(1999){Arnaud}, {Dorman}, \& {Gordon}}]{XSPEC1999}
{Arnaud}, K., {Dorman}, B., \& {Gordon}, C. 1999, {XSPEC: An X-ray spectral
  fitting package}, , , ascl:9910.005

\bibitem[{{Axelsson} {et~al.}(2012){Axelsson}, {Baldini}, \&
  et~al.}]{Axelsson2012}
{Axelsson}, M., {Baldini}, L., \& et~al. 2012, ApJ, 757, L31

\bibitem[{{Axelsson} \& {Borgonovo}(2015)}]{Axelsson2015}
{Axelsson}, M., \& {Borgonovo}, L. 2015, MNRAS, 447, 3150

\bibitem[{{Band} {et~al.}(1993){Band}, {Matteson}, \& et~al.}]{Band1993}
{Band}, D., {Matteson}, J., \& et~al. 1993, ApJ, 413, 281

\bibitem[{{Baring} \& {Braby}(2004)}]{BaringBraby2004}
{Baring}, M.~G., \& {Braby}, M.~L. 2004, The Astrophysical Journal, 613, 460

\bibitem[{B{\'e}gu{\'e} {et~al.}(2013)B{\'e}gu{\'e}, Siutsou, \&
  Vereshchagin}]{begue2013monte}
B{\'e}gu{\'e}, D., Siutsou, I., \& Vereshchagin, G. 2013, The Astrophysical
  Journal, 767, 139

\bibitem[{{Beloborodov}(2010)}]{Beloborodov2010}
{Beloborodov}, A.~M. 2010, MNRAS, 407, 1033

\bibitem[{{Beloborodov}(2011)}]{Beloborodov2011}
---. 2011, ApJ, 737, 68

\bibitem[{{Beloborodov} {et~al.}(2014){Beloborodov}, {Hasco{\"e}t}, \&
  {Vurm}}]{Beloborodov2014}
{Beloborodov}, A.~M., {Hasco{\"e}t}, R., \& {Vurm}, I. 2014, \apj, 788, 36

\bibitem[{{Beniamini} \& {Piran}(2013)}]{Beniamini&Piran2013}
{Beniamini}, P., \& {Piran}, T. 2013, ApJ, 769, 69

\bibitem[{{Bhattacharya} \& {Kumar}(2019)}]{BhattacharyaKumar2019}
{Bhattacharya}, M., \& {Kumar}, P. 2019, arXiv e-prints, arXiv:1909.07398

\bibitem[{{Burgess}(2019)}]{Burgess2019_Width}
{Burgess}, J.~M. 2019, \aap, 629, A69

\bibitem[{{Burgess} {et~al.}(2018){Burgess}, {B{\'e}gu{\'e}}, {Bacelj},
  {Giannios}, {Berlato}, \& {Greiner}}]{Burgess2018}
{Burgess}, J.~M., {B{\'e}gu{\'e}}, D., {Bacelj}, A., {et~al.} 2018, arXiv
  e-prints, arXiv:1810.06965

\bibitem[{{Burgess} {et~al.}(2016){Burgess}, {B{\'e}gu{\'e}}, {Ryde}, {Omodei},
  {Pe'er}, {Racusin}, \& {Cucchiara}}]{Burgess2016}
{Burgess}, J.~M., {B{\'e}gu{\'e}}, D., {Ryde}, F., {et~al.} 2016, \apj, 822, 63

\bibitem[{{Burgess} {et~al.}(2019{\natexlab{a}}){Burgess}, {Kole}, {Berlato},
  {Greiner}, {Vianello}, {Produit}, {Li}, \& {Sun}}]{Burgess2019}
{Burgess}, J.~M., {Kole}, M., {Berlato}, F., {et~al.} 2019{\natexlab{a}}, \aap,
  627, A105

\bibitem[{{Burgess} {et~al.}(2019{\natexlab{b}}){Burgess}, {Kole}, {Berlato},
  {Greiner}, {Vianello}, {Produit}, {Li}, \& {Sun}}]{Burgess2019_170114}
---. 2019{\natexlab{b}}, \aap, 627, A105

\bibitem[{{Burgess} {et~al.}(2011){Burgess}, {Preece}, \& et~al.}]{Burgess2011}
{Burgess}, J.~M., {Preece}, R.~D., \& et~al. 2011, ApJ, 741, 24

\bibitem[{{Burgess} {et~al.}(2015){Burgess}, {Ryde}, \&
  {Yu}}]{Burgess2015_alpha}
{Burgess}, J.~M., {Ryde}, F., \& {Yu}, H.-F. 2015, \mnras, 451, 1511

\bibitem[{{Chhotray} \& {Lazzati}(2018)}]{Chhotray2018}
{Chhotray}, A., \& {Lazzati}, D. 2018, \mnras, 476, 2352

\bibitem[{{Crider} {et~al.}(1997){Crider}, {Liang}, {Smith}, {Preece},
  {Briggs}, {Pendleton}, {Paciesas}, {Band}, \& {Matteson}}]{Crider1997}
{Crider}, A., {Liang}, E.~P., {Smith}, I.~A., {et~al.} 1997, ApJL, 479, L39

\bibitem[{{Derishev} \& {Piran}(2019)}]{DerishevPiran2019}
{Derishev}, E., \& {Piran}, T. 2019, \apjl, 880, L27

\bibitem[{{Epstein} \& {Petrosian}(1973)}]{Epstein1973}
{Epstein}, R.~I., \& {Petrosian}, V. 1973, The Astrophysical Journal, 183, 611

\bibitem[{{Feroz} \& {Hobson}(2008)}]{ferozhobson2008}
{Feroz}, F., \& {Hobson}, M.~P. 2008, \mnras, 384, 449

\bibitem[{{Feroz} {et~al.}(2011){Feroz}, {Hobson}, \&
  {Bridges}}]{ferozhobsonbridges}
{Feroz}, F., {Hobson}, M.~P., \& {Bridges}, M. 2011, {MultiNest: Efficient and
  Robust Bayesian Inference}, , , ascl:1109.006

\bibitem[{{Fraija} {et~al.}(2019){Fraija}, {Barniol Duran}, {Dichiara}, \&
  {Beniamini}}]{Fraija2019}
{Fraija}, N., {Barniol Duran}, R., {Dichiara}, S., \& {Beniamini}, P. 2019,
  arXiv e-prints, arXiv:1907.06675

\bibitem[{{Fraija} {et~al.}(2017){Fraija}, {Lee}, {Araya}, {Veres}, {Barniol
  Duran}, \& {Guiriec}}]{Fraija2017}
{Fraija}, N., {Lee}, W.~H., {Araya}, M., {et~al.} 2017, \apj, 848, 94

\bibitem[{{Gabry} {et~al.}(2017){Gabry}, {Simpson}, {Vehtari}, {Betancourt}, \&
  {Gelman}}]{gelmanvis}
{Gabry}, J., {Simpson}, D., {Vehtari}, A., {Betancourt}, M., \& {Gelman}, A.
  2017, arXiv e-prints, arXiv:1709.01449

\bibitem[{Gelman(2004)}]{gelman_2004}
Gelman, A. 2004, Journal of Computational and Graphical Statistics, 13,
  755–779

\bibitem[{Gelman {et~al.}(2014)Gelman, Hwang, \& Vehtari}]{Gelman2014}
Gelman, A., Hwang, J., \& Vehtari, A. 2014, Statistics and Computing, 24, 997.
\newblock \url{http://dx.doi.org/10.1007/s11222-013-9416-2}

\bibitem[{{Ghirlanda} {et~al.}(2010){Ghirlanda}, {Ghisellini}, \&
  {Nava}}]{Ghirlanda2010}
{Ghirlanda}, G., {Ghisellini}, G., \& {Nava}, L. 2010, \aap, 510, L7

\bibitem[{{Ghisellini} {et~al.}(2019){Ghisellini}, {Ghirlanda}, {Oganesyan},
  {Ascenzi}, {Nava}, {Celotti}, {Salafia}, {Ravasio}, \&
  {Ronchi}}]{Ghisellini2019}
{Ghisellini}, G., {Ghirlanda}, G., {Oganesyan}, G., {et~al.} 2019, arXiv
  e-prints, arXiv:1912.02185

\bibitem[{{Giannios}(2006)}]{Giannios2006}
{Giannios}, D. 2006, A{$\&$}A, 457, 763

\bibitem[{{Giuliani} {et~al.}(2014){Giuliani}, {Mereghetti}, {Marisaldi},
  {Longo}, {Del Monte}, {Pittori}, {Verrecchia}, {Tavani}, {Cattaneo},
  {Pacciani}, {Vercellone}, \& {Rappoldi}}]{Giuliani2014}
{Giuliani}, A., {Mereghetti}, S., {Marisaldi}, M., {et~al.} 2014, arXiv
  e-prints, arXiv:1407.0238

\bibitem[{{Goldstein} {et~al.}(2012){Goldstein}, {Burgess}, {Preece}, {Briggs},
  {Guiriec}, {van der Horst}, {Connaughton}, {Wilson-Hodge}, {Paciesas},
  {Meegan}, {von Kienlin}, {Bhat}, {Bissaldi}, {Chaplin}, {Diehl}, {Fishman},
  {Fitzpatrick}, {Foley}, {Gibby}, {Giles}, {Greiner}, {Gruber}, {Kippen},
  {Kouveliotou}, {McBreen}, {McGlynn}, {Rau}, \& {Tierney}}]{Goldstein2012}
{Goldstein}, A., {Burgess}, J.~M., {Preece}, R.~D., {et~al.} 2012, \apjs, 199,
  19

\bibitem[{{Golenetskii} {et~al.}(1983){Golenetskii}, {Mazets}, {Aptekar}, \&
  {Ilinskii}}]{Golenetskii1983}
{Golenetskii}, S.~V., {Mazets}, E.~P., {Aptekar}, R.~L., \& {Ilinskii}, V.~N.
  1983, \nat, 306, 451

\bibitem[{{Golkhou} \& {Butler}(2014)}]{Golkhou&Butler2014}
{Golkhou}, V.~Z., \& {Butler}, N.~R. 2014, ApJ, 787, 90

\bibitem[{{Guiriec} {et~al.}(2015){Guiriec}, {Kouveliotou}, {Daigne}, {Zhang},
  {Hasco{\"e}t}, {Nemmen}, {Thompson}, {Bhat}, {Gehrels}, {Gonzalez}, {Kaneko},
  {McEnery}, {Mochkovitch}, {Racusin}, {Ryde}, {Sacahui}, \&
  {{\"U}nsal}}]{Guiriec2015_New_Model}
{Guiriec}, S., {Kouveliotou}, C., {Daigne}, F., {et~al.} 2015, \apj, 807, 148

\bibitem[{{Ito} {et~al.}(2013){Ito}, {Nagataki}, {Ono}, {Lee}, {Mao}, {Yamada},
  {Pe'er}, {Mizuta}, \& {Harikae}}]{Ito2013}
{Ito}, H., {Nagataki}, S., {Ono}, M., {et~al.} 2013, \apj, 777, 62

\bibitem[{{Iyyani} {et~al.}(2016){Iyyani}, {Ryde}, {Burgess}, {Pe'er}, \&
  {B{\'e}gu{\'e}}}]{Iyyani2016}
{Iyyani}, S., {Ryde}, F., {Burgess}, J.~M., {Pe'er}, A., \& {B{\'e}gu{\'e}}, D.
  2016, \mnras, 456, 2157

\bibitem[{{Jaynes} \& {Bretthorst}(2003)}]{jaynesbook}
{Jaynes}, E.~T., \& {Bretthorst}, G.~L. 2003, {Probability Theory}

\bibitem[{{Jeffreys}(1957)}]{jeffreysastropaper}
{Jeffreys}, H. 1957, \mnras, 117, 347

\bibitem[{Jeffreys(1961)}]{Jeffreysbook}
Jeffreys, H. 1961, Theory of Probability, 3rd edn. (Oxford, England: Oxford)

\bibitem[{{Kargatis} {et~al.}(1994){Kargatis}, {Liang}, {Hurley}, {Barat},
  {Eveno}, \& {Niel}}]{Kargatis1994}
{Kargatis}, V.~E., {Liang}, E.~P., {Hurley}, K.~C., {et~al.} 1994, \apj, 422,
  260

\bibitem[{Kass \& Raftery(1995)}]{kassraftery}
Kass, R.~E., \& Raftery, A.~E. 1995, Journal of the American Statistical
  Association, 90, 773.
\newblock
  \url{https://amstat.tandfonline.com/doi/abs/10.1080/01621459.1995.10476572}

\bibitem[{Kass \& Wasserman(1995)}]{bic}
Kass, R.~E., \& Wasserman, L. 1995, Journal of the American Statistical
  Association, 90, 928.
\newblock
  \url{https://www.tandfonline.com/doi/abs/10.1080/01621459.1995.10476592}

\bibitem[{{Kumar} \& {Barniol Duran}(2009)}]{Kumar2009}
{Kumar}, P., \& {Barniol Duran}, R. 2009, \mnras, 400, L75

\bibitem[{{Li}(2019)}]{Li2019BBPointing}
{Li}, L. 2019, \apjs, 242, 16

\bibitem[{{Lloyd} \& {Petrosian}(1999)}]{Lloyd1999}
{Lloyd}, N.~M., \& {Petrosian}, V. 1999, \apj, 511, 550

\bibitem[{{Lloyd} \& {Petrosian}(2000)}]{Lloyd2000}
---. 2000, \apj, 543, 722

\bibitem[{{Lloyd-Ronning} \& {Petrosian}(2002)}]{Lloyd2002}
{Lloyd-Ronning}, N.~M., \& {Petrosian}, V. 2002, \apj, 565, 182

\bibitem[{{Lundman} {et~al.}(2013){Lundman}, {Pe'er}, \& {Ryde}}]{Lundman2013}
{Lundman}, C., {Pe'er}, A., \& {Ryde}, F. 2013, MNRAS, 428, 2430

\bibitem[{{Medvedev}(2000)}]{Medvedev2000}
{Medvedev}, M.~V. 2000, \apj, 540, 704

\bibitem[{{Meng} {et~al.}(2019){Meng}, {Liu}, {Wei}, {Wu}, \&
  {Zhang}}]{Meng2019}
{Meng}, Y.-Z., {Liu}, L.-D., {Wei}, J.-J., {Wu}, X.-F., \& {Zhang}, B.-B. 2019,
  arXiv e-prints, arXiv:1904.08526

\bibitem[{{M{\'e}sz{\'a}ros}(2019)}]{Meszaros2019}
{M{\'e}sz{\'a}ros}, P. 2019, arXiv e-prints, arXiv:1904.10488

\bibitem[{{M{\'e}sz{\'a}ros} {et~al.}(2002){M{\'e}sz{\'a}ros}, {Ramirez-Ruiz},
  \& et~al.}]{Meszaros2002}
{M{\'e}sz{\'a}ros}, P., {Ramirez-Ruiz}, E., \& et~al. 2002, ApJ, 578, 812

\bibitem[{Nakar \& Piran(2017)}]{Nakar2017}
Nakar, E., \& Piran, T. 2017, Astrophysical Journal, 834, 28

\bibitem[{{Oganesyan} {et~al.}(2017){Oganesyan}, {Nava}, {Ghirlanda}, \&
  {Celotti}}]{Oganesyan2018}
{Oganesyan}, G., {Nava}, L., {Ghirlanda}, G., \& {Celotti}, A. 2017, ArXiv
  e-prints, arXiv:1710.09383

\bibitem[{{Oganesyan} {et~al.}(2019){Oganesyan}, {Nava}, {Ghirlanda},
  {Melandri}, \& {Celotti}}]{Oganesyan2019}
{Oganesyan}, G., {Nava}, L., {Ghirlanda}, G., {Melandri}, A., \& {Celotti}, A.
  2019, arXiv e-prints, arXiv:1904.11086

\bibitem[{{Panaitescu}(2017)}]{Panaitescu2017}
{Panaitescu}, A. 2017, \apj, 837, 13

\bibitem[{{Panaitescu} \& {M{\'e}sz{\'a}ros}(1998)}]{Panaitescu1998}
{Panaitescu}, A., \& {M{\'e}sz{\'a}ros}, P. 1998, ApJ, 492, 683

\bibitem[{{Pe'er} {et~al.}(2015){Pe'er}, {Barlow}, {O'Mahony}, {Margutti},
  {Ryde}, {Larsson}, {Lazzati}, \& {Livio}}]{peer2015}
{Pe'er}, A., {Barlow}, H., {O'Mahony}, S., {et~al.} 2015, \apj, 813, 127

\bibitem[{{Pe'er} {et~al.}(2005){Pe'er}, {M{\'e}sz{\'a}ros}, \&
  {Rees}}]{Peer2005}
{Pe'er}, A., {M{\'e}sz{\'a}ros}, P., \& {Rees}, M.~J. 2005, ApJ, 635, 476

\bibitem[{{Pe'er} {et~al.}(2006){Pe'er}, {M{\'e}sz{\'a}ros}, \&
  {Rees}}]{Peer2006}
---. 2006, ApJ, 642, 995

\bibitem[{{Preece} {et~al.}(1998){Preece}, {Briggs}, {Mallozzi}, {Pendleton},
  {Paciesas}, \& {Band}}]{Preece1998}
{Preece}, R.~D., {Briggs}, M.~S., {Mallozzi}, R.~S., {et~al.} 1998, ApJL, 506,
  L23

\bibitem[{{Ravasio} {et~al.}(2019){Ravasio}, {Oganesyan}, {Salafia}, {Ghirland
  a}, {Ghisellini}, {Branchesi}, {Campana}, {Covino}, \&
  {Salvaterra}}]{Ravasio2019}
{Ravasio}, M.~E., {Oganesyan}, G., {Salafia}, O.~S., {et~al.} 2019, \aap, 626,
  A12

\bibitem[{{Rees} \& {M\'esz\'aros}(1992)}]{ReesMeszaros1992}
{Rees}, M.~J., \& {M\'esz\'aros}, P. 1992, \mnras, 258, 41

\bibitem[{{Rees} \& {M{\'e}sz{\'a}ros}(1994)}]{Rees1994}
{Rees}, M.~J., \& {M{\'e}sz{\'a}ros}, P. 1994, ApJL, 430, L93

\bibitem[{{Rees} \& {M{\'e}sz{\'a}ros}(2005)}]{Rees&Meszaros2005}
---. 2005, ApJ, 628, 847

\bibitem[{{Ronchi} {et~al.}(2019){Ronchi}, {Fumagalli}, {Ravasio}, {Oganesyan},
  {Toffano}, {Salafia}, {Nava}, {Ascenzi}, {Ghirlanda}, \&
  {Ghisellini}}]{Ronchi2019}
{Ronchi}, M., {Fumagalli}, F., {Ravasio}, M.~E., {et~al.} 2019, arXiv e-prints,
  arXiv:1909.10531

\bibitem[{{Ryde}(2005)}]{Ryde2005}
{Ryde}, F. 2005, ApJ, 625, L95

\bibitem[{{Ryde} {et~al.}(2017){Ryde}, {Lundman}, \& {Acuner}}]{Ryde2017}
{Ryde}, F., {Lundman}, C., \& {Acuner}, Z. 2017, \mnras, 472, 1897

\bibitem[{{Ryde} {et~al.}(2011){Ryde}, {Pe'er}, \& et~al.}]{Ryde2011}
{Ryde}, F., {Pe'er}, A., \& et~al. 2011, MNRAS, 415, 3693

\bibitem[{{Ryde} {et~al.}(2019){Ryde}, {Yu}, {Dereli-B{\'e}gu{\'e}}, {Lundman},
  {Pe'er}, \& {Li}}]{Ryde2019}
{Ryde}, F., {Yu}, H.-F., {Dereli-B{\'e}gu{\'e}}, H., {et~al.} 2019, \mnras,
  arXiv:1901.01775

\bibitem[{{Sharma} {et~al.}(2020){Sharma}, {Iyyani}, {Bhattacharya},
  {Chattopadhyay}, {Vadawale}, \& {Bhalerao}}]{Sharma2020}
{Sharma}, V., {Iyyani}, S., {Bhattacharya}, D., {et~al.} 2020, arXiv e-prints,
  arXiv:2003.02284

\bibitem[{{Sharma} {et~al.}(2019){Sharma}, {Iyyani}, {Bhattacharya},
  {Chattopadhyay}, {Rao}, {Aarthy}, {Vadawale}, {Mithun}, {Bhalerao}, {Ryde},
  \& {Peer}}]{Sharma2019}
---. 2019, arXiv e-prints, arXiv:1908.10885

\bibitem[{{Sironi} {et~al.}(2015){Sironi}, {Keshet}, \& {Lemoine}}]{Sironi2015}
{Sironi}, L., {Keshet}, U., \& {Lemoine}, M. 2015, \ssr, 191, 519

\bibitem[{{Skilling}(2004)}]{skilling}
{Skilling}, J. 2004, in American Institute of Physics Conference Series, ed.
  R.~{Fischer}, R.~{Preuss}, \& U.~V. {Toussaint}, Vol. 735, 395--405

\bibitem[{{Tavani}(1996)}]{Tavani1996}
{Tavani}, M. 1996, ApJ, 466, 768

\bibitem[{{Thompson} \& {Gill}(2014)}]{TompsonGill2014}
{Thompson}, C., \& {Gill}, R. 2014, \apj, 791, 46

\bibitem[{{Vianello}(2018)}]{Vianello2018}
{Vianello}, G. 2018, \apjs, 236, 17

\bibitem[{{Vurm} \& {Beloborodov}(2016)}]{Vurm2016}
{Vurm}, I., \& {Beloborodov}, A.~M. 2016, \apj, 831, 175

\bibitem[{{Vurm} {et~al.}(2014){Vurm}, {Hasco{\"e}t}, \&
  {Beloborodov}}]{Vurm2014}
{Vurm}, I., {Hasco{\"e}t}, R., \& {Beloborodov}, A.~M. 2014, \apjl, 789, L37

\bibitem[{{Wheaton} {et~al.}(1973){Wheaton}, {Ulmer}, {Baity}, {Datlowe},
  {Elcan}, {Peterson}, {Klebesadel}, {Strong}, {Cline}, \&
  {Desai}}]{Weaton1973}
{Wheaton}, W.~A., {Ulmer}, M.~P., {Baity}, W.~A., {et~al.} 1973, The
  Astrophysical Journal, 185, L57

\bibitem[{{Wijers} \& {Galama}(1999)}]{WijersGalama1999}
{Wijers}, R.~A.~M.~J., \& {Galama}, T.~J. 1999, \apj, 523, 177

\bibitem[{{Yang} \& {Zhang}(2018)}]{Yang2018}
{Yang}, Y.-P., \& {Zhang}, B. 2018, The Astrophysical Journal, 864, L16

\bibitem[{{Yassine} {et~al.}(2017){Yassine}, {Piron}, {Mochkovitch}, \&
  {Daigne}}]{Yassine2017}
{Yassine}, M., {Piron}, F., {Mochkovitch}, R., \& {Daigne}, F. 2017, \aap, 606,
  A93

\bibitem[{{Yu} {et~al.}(2019){Yu}, {Dereli-B{\'e}gu{\'e}}, \& {Ryde}}]{Yu2019}
{Yu}, H.-F., {Dereli-B{\'e}gu{\'e}}, H., \& {Ryde}, F. 2019, arXiv e-prints,
  arXiv:1810.07313

\bibitem[{{Yu} {et~al.}(2016){Yu}, {Preece}, {Greiner}, {Narayana Bhat},
  {Bissaldi}, {Briggs}, {Cleveland}, {Connaughton}, {Goldstein}, {von Kienlin},
  {Kouveliotou}, {Mailyan}, {Meegan}, {Paciesas}, {Rau}, {Roberts}, {Veres},
  {Wilson-Hodge}, {Zhang}, \& {van Eerten}}]{Yu2016}
{Yu}, H.-F., {Preece}, R.~D., {Greiner}, J., {et~al.} 2016, \aap, 588, A135

\bibitem[{{Zabalza}(2015)}]{NaimaPackage}
{Zabalza}, V. 2015, Proc.~of International Cosmic Ray Conference 2015, 922

\bibitem[{{Zhang} {et~al.}(2018){Zhang}, {Zhang}, {Castro-Tirado}, {Dai},
  {Tam}, {Wang}, {Hu}, {Karpov}, {Pozanenko}, {Zhang}, {Mazaeva}, {Minaev},
  {Volnova}, {Oates}, {Gao}, {Wu}, {Shao}, {Tang}, {Beskin}, {Biryukov},
  {Bondar}, {Ivanov}, {Katkova}, {Orekhova}, {Perkov}, {Sasyuk}, {Mankiewicz},
  {{\.Z}arnecki}, {Cwiek}, {Opiela}, {Zadro{\.Z}ny}, {Aptekar}, {Frederiks},
  {Svinkin}, {Kusakin}, {Inasaridze}, {Burhonov}, {Rumyantsev}, {Klunko},
  {Moskvitin}, {Fatkhullin}, {Sokolov}, {Valeev}, {Jeong}, {Park},
  {Caballero-Garc{\'{\i}}a}, {Cunniffe}, {Tello}, {Ferrero}, {Pandey},
  {Jel{\'{\i}}nek}, {Peng}, {S{\'a}nchez-Ram{\'{\i}}rez}, \&
  {Castell{\'o}n}}]{ZhangBB2018}
{Zhang}, B.-B., {Zhang}, B., {Castro-Tirado}, A.~J., {et~al.} 2018, Nature
  Astronomy, 2, 69

\end{thebibliography}



\end{document}